\documentclass[prb,twocolumn,superscriptaddress]{revtex4}

\usepackage{graphicx}
\usepackage{txfonts}
\usepackage{bm}
\usepackage{color}
\usepackage{hyperref}

\begin{document}

\title{Observation of persistent centrosymmetricity in the hexagonal manganite family}

\author{Yu Kumagai}
\email[]{yu.kumagai@mat.ethz.ch}
\affiliation{Department of Materials, ETH Zurich, Wolfgang-Pauli-Strasse 10, 8093 Zurich, Switzerland}
\author{Alexei A. Belik}
\affiliation{International Center for Materials Nanoarchitectonics (WPI-MANA), NIMS, Namiki 1-1, Ibaraki 305-0044, Japan}
\author{Martin Lilienblum}
\affiliation{Department of Materials, ETH Zurich, Wolfgang-Pauli-Strasse 10, 8093 Zurich, Switzerland}
\author{Na$\ddot{\rm e}$mi Leo}
\affiliation{Department of Materials, ETH Zurich, Wolfgang-Pauli-Strasse 10, 8093 Zurich, Switzerland}
\author{Manfred Fiebig}
\affiliation{Department of Materials, ETH Zurich, Wolfgang-Pauli-Strasse 10, 8093 Zurich, Switzerland}
\author{Nicola A. Spaldin}
\affiliation{Department of Materials, ETH Zurich, Wolfgang-Pauli-Strasse 10, 8093 Zurich, Switzerland}
\date{\today}

\definecolor{blue}{rgb}{0.0,0.0,1}
\hypersetup{colorlinks,breaklinks,linkcolor=blue,urlcolor=blue,anchorcolor=blue,citecolor=blue}

\begin{abstract}
The controversy regarding the ferroelectric behavior of hexagonal InMnO$_3$ is resolved by using a combination 
of x-ray diffraction (XRD), piezoresponse force microscopy (PFM), second harmonic generation (SHG), and density functional theory (DFT). 
While XRD data show a symmetry-lowering unit-cell tripling,  which is also found in the multiferroic hexagonal manganites of $P6_3cm$ symmetry,
PFM and SHG do not detect ferroelectricity at ambient or low temperature, in striking contrast to the behavior in the multiferroic
counterparts. We propose instead a centrosymmetric $P\overline{3}c$ phase as the ground state structure.
Our DFT calculations reveal that the relative energy of the ferroelectric and
nonferroelectric structures is determined by a competition between electrostatics and oxygen-$R$-site covalency, with
an $absence$ of covalency favoring the ferroelectric phase. 
\end{abstract}

% 75.85.+t Magnetoelectric effects, multiferroics
% 77.80.-e Ferroelectricity and antiferroelectricity

% 77.84.Bw Dielectric, piezoelectric, ferroelectric, and antiferroelectric materials: Elements, oxides, nitrides, borides, carbides, chalcogenides, etc.
% 61.66.Fn Structure of specific crystalline solids: Inorganic compounds
% 71.15.Nc Total energy and cohesive energy calculations
% 71.20.-b Electron density of states and band structure of crystalline solids

% 71.15.Mb Density functional theory, local density approximation, gradient and other corrections

\maketitle
\section{Introduction: structure of the hexagonal manganites}
\begin{figure}
  \includegraphics[width=1\linewidth]{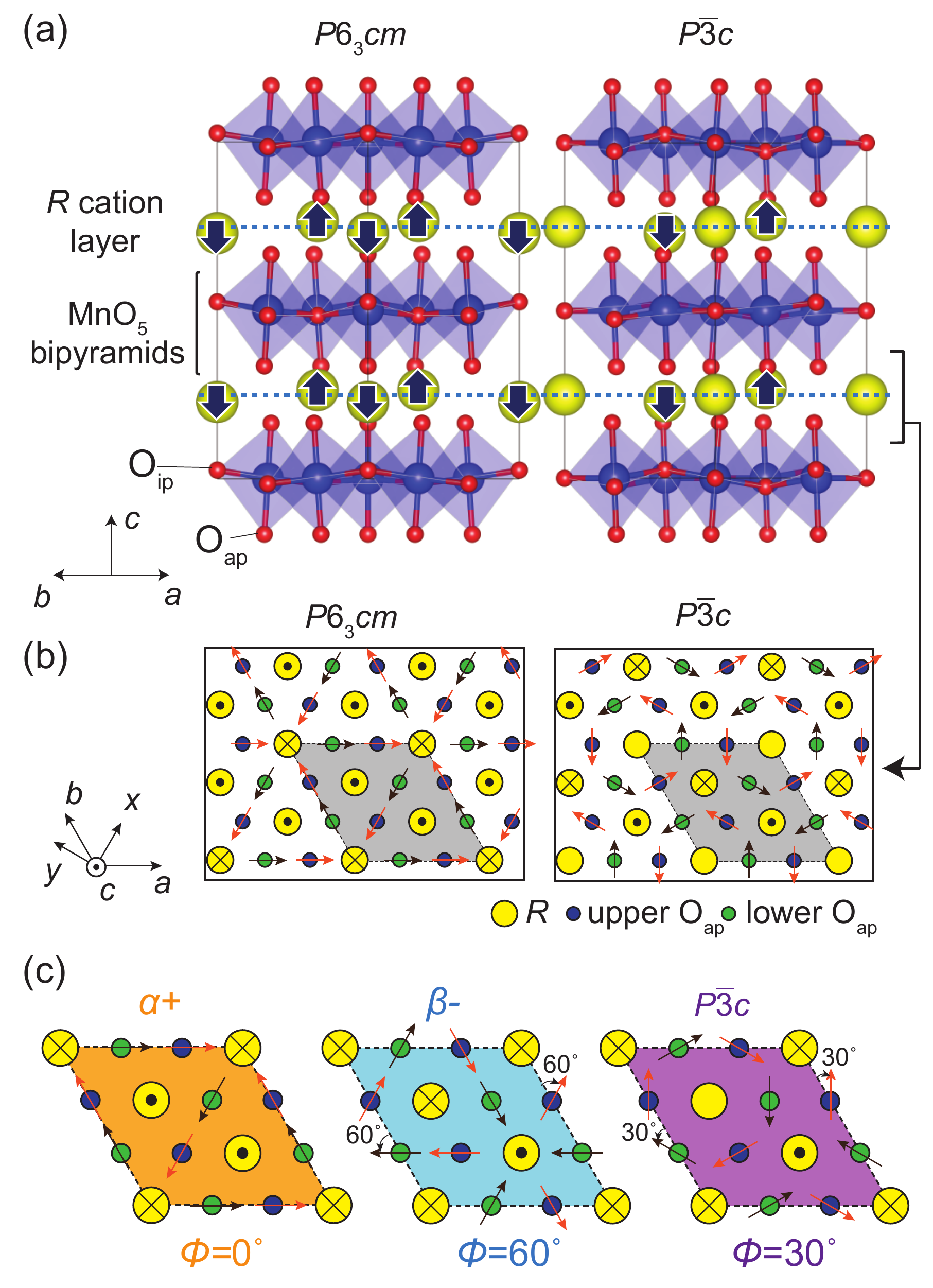}
  \caption{(a) Side and (b) top views of the two candidate InMnO$_3$ structures. 
           $P6_3cm$ is the ferroelectric phase of the h-$R$MnO$_3$ compounds; $P\overline{3}c$ is the structure proposed for InMnO$_3$ in this study.
           Arrows indicate the displacements from the high symmetry $P6_3/mmc$ phase.
           The primitive unit cell is shaded grey.
           In contrast with the $P6_3cm$ phase, one In ion in the $P\overline{3}c$ unit cell remains at the high-symmetry $2b$ site, retaining the inversion symmetry. 
           Note also the different tilt patterns of the MnO$_5$ polyhedra toward or around corner $R$ ions.
           (c) Top views of $\alpha^+$ and $\beta^-$ domains in the $P6_3cm$ phase using the notation from Ref.~\onlinecite{NatMat.9.253} and one of six domains in the $P\overline{3}c$ phase.
           The phases ${\it \Phi}$, defined by the (counter)clockwise angle of tilting direction of upper (lower) oxygen layers relative to $\alpha^+$ domain, are also shown (Ref.~\onlinecite{NatMat.9.253}).
Note that the $P\overline{3}c$ phase has a tilt phase of 30$^{\circ}$+$n\cdot60^{\circ}$, and is obtained by averaging the tilt patterns and
$R$-ion displacements of two $P6_3cm$ trimerization domains with different origins and orientations 
such as the $\alpha^+$ and $\beta^-$ domains.
  }
  
  \label{str}
\end{figure}

Hexagonal h-$R$MnO$_3$ ($R$ = Sc, Y, Dy--Lu) represents an established class of multiferroics in which ferroelectricity and antiferromagnetism exist simultaneously.
Although their fundamental properties have been investigated for half a century, recent reports of intriguing characteristics 
such as interlocked antiphase (AP) and ferroelectric (FE) (AP+FE) domain walls\cite{NatMat.9.253, ApplPhysLett.97.012904,NatMat.submitted} are fueling continued interest.
At the root of these behaviors is their unusual improper geometric ferroelectricity,\cite{NatMat.430.541,PhysRevB.72.100103} which is
in turn related to their layered structure, in which $xy$ planes of $R^{3+}$ ions are interspaced by layers of corner-shared 
MnO$_5$ trigonal polyhedra (Fig.~\ref{str}).

%==========================================================================================================
%==========================================================================================================

InMnO$_3$ crystallizes in the same hexagonal manganite structure as the h-$R$MnO$_3$ compounds, and might be expected to show analogous ferroelectric behavior. 
In fact, most previous x-ray and neutron powder diffraction refinements assigned InMnO$_3$ to the polar $P6_3cm$ structure adopted by
the multiferroic hexagonal manganites.\cite{JSolidStateChem.116.118,PhysRevB.79.054411,PhysRevB.84.054455}
In the $P6_3cm$ structure, the MnO$_5$ trigonal bipyramids tilt and trimerize with a trimerization phase of $n\cdot60^{\circ}$, where $n$ is an integer.
The $R$ ions on the $2a$ sites displace up or down along the $z$ direction, depending on the tilting direction, and those on the $4b$ sites in the opposite direction (Fig.~\ref{str}).
This tilt symmetry then enables an additional displacement of the $R$ sublattice relative to the Mn-O layers causing a net ferroelectric polarization.
Rusakov and Belik $et~al.$ pointed out that the nonpolar $P\overline{3}c$ structure -- in which the MnO$_5$ polyhedra trimerize
at intermediate angles and the inversion symmetry is retained (Fig.~\ref{str}) -- and polar $P6_3cm$ structure have similar 
powder x-ray-diffraction $R$ values.\cite{InorgChem.50.3559}
They disregarded the $P\overline{3}c$ model in their subsequent analysis, however, believing that all h-$R$MnO$_3$ compounds should be polar.
Indeed, ferroelectricity has been reported in InMnO$_3$ below 500~K based on the observation of polarization-electric field (P-E) hysteresis loops obtained by a ferroelectric test system.
Such pyroelectric current measurements, especially if they are applied to amorphous samples or thin films, are notoriously sensitive to sample defects, however,
and the P-E loops shown in Ref.~\onlinecite{JApplPhys.100.076104} could indicate leaky dielectric behavior rather than ferroelectricity.\cite{JPhysCondensMatter.20.021001}
In agreement with this, Belik $et~al.$ did not observe spontaneous polarization when they repeated the experiment.\cite{PhysRevB.79.054411,InorgChem.50.3559}
Therefore there is no clear evidence to date that InMnO$_3$ has the $P6_3cm$ structure or shows ferroelectric polarization.

In this study, we revisit the structure and polarization behavior of InMnO$_3$ 
by using a combination of x-ray diffraction (XRD), piezoresponse force microscopy (PFM), optical second harmonic generation (SHG), 
and density functional theory (DFT) and show that InMnO$_3$ is indeed centrosymmetric, with $P\overline{3}c$ as the most likely space group (Fig.~\ref{str}).
We explain the difference between InMnO$_3$ and the multiferroic h-$R$MnO$_3$ compounds using DFT analysis of
the chemical bonding, and propose another candidate material TlMnO$_3$ that should also show the nonferroelectric InMnO$_3$ structure.
Finally, we discuss the implications of this newly-identified structure for the multiferroicity in the h-$R$MnO$_3$ family in general.

\section{Experiments}

%==========================================================================================================
\begin{figure}
  \includegraphics[width=1\linewidth]{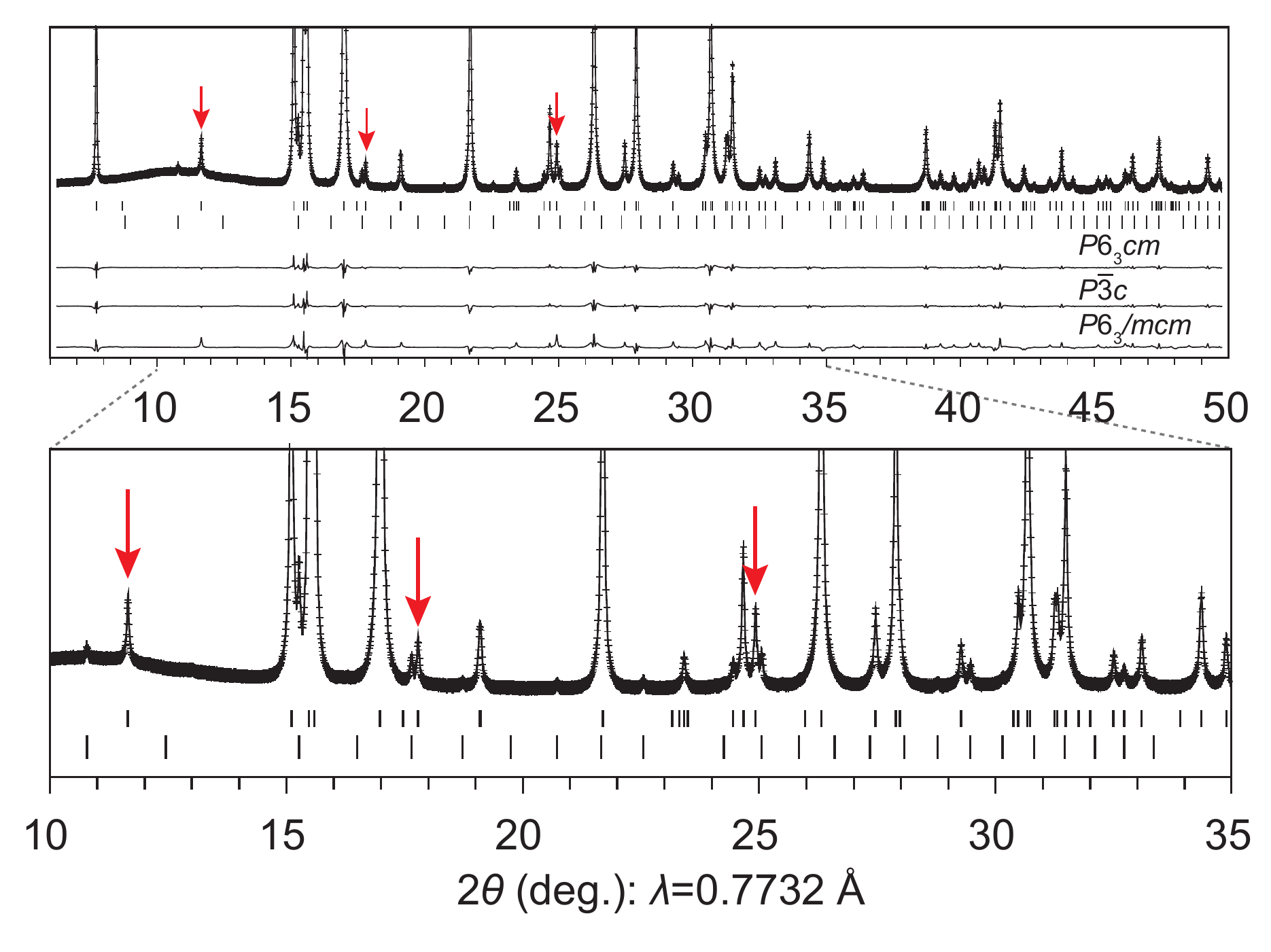}
  \caption{Synchrotron x-ray powder diffraction patterns of InMnO$_3$ at 293~K. 
           Crosses represent data points and solid lines calculated intensities for $P\overline{3}c$.
           Differences for three structure models are also shown.
           Bragg reflections are indicated by tick marks (these are the same for $P6_3cm$, $P\overline{3}c$, and $P6_3/mcm$ models). 
           The lower tick marks indicate reflections from In$_2$O$_3$ impurity. 
           Arrows show reflections corresponding to unit-cell tripling.
}
  \label{xrd}
\end{figure}
%==========================================================================================================

%==========================================================================================================
\begin{table}
 \caption{Atomic fractional coordinates for InMnO$_3$ at 293~K in space group $P\overline{3}c$ (top) and $P6_3cm$ (bottom), obtained in this work using powder XRD.
          Our measured lattice constants $a$ and $c$ are 5.88462(10) and 11.48540(15)~$\AA$ for $P\overline{3}c$ and 5.88463(7) and 11.48541(12)~$\AA$ for $P6_3cm$, respectively.
}
\label{XRDTable}
  \begin{center}
  \begin{tabular}{cccccc}\\ 
  \hline
  \hline
    \multicolumn{6}{c}{$P\overline{3}c$}     \\
  \hline
  Atom  & Wyckoff position & $x$ & $y$ & $z$ & $B_{\rm iso}$ ($\AA^2$)   \\
  \hline
     In1 & $4d$ & 1/3 & 2/3 & 0.51674(8) & 0.46(3) \\
     In2 & $2b$ & 0   & 0   & 0          & 1.28(8) \\
     Mn  & $6f$ & 0.6587(10) & 0 & 1/4   & 0.36(2) \\
     O1  & $2a$ & 0   & 0   & 1/4        & 0.9(4) \\
     O2  & $4d$ & 1/3 & 2/3 & 0.7312(7)  & 0.26(15)\\
     O3  & $12g$ & 0.6829(25) & 0.0241(10) & 0.0858(2) & 0.77(8) \\
  \hline
  \hline
  \end{tabular}
  \begin{tabular}{cccccc}\\ 
  \hline
  \hline
    \multicolumn{6}{c}{$P6_3cm$}     \\
  \hline
  Atom  & Wyckoff position & $x$ & $y$ & $z$ & $B_{\rm iso}$ ($\AA^2$)   \\
  \hline
     In1 & $2a$ & 0 & 0 & 0.2674(6) & 0.33(6) \\
     In2 & $4b$ & 1/3 & 2/3  & 0.2383(6) & 0.92(4) \\
     Mn  & $6c$ & 0.3250(10) & 0 & 0   & 0.37(3) \\
     O1  & $6c$ & 0.3117(22) & 0 & 0.1749(11) & 1.5(3) \\
     O2  & $6c$ & 0.6466(18) & 0 & 0.3445(10) & -0.4(2)\\
     O3  & $2a$ & 0 & 0 & 0.4746(20) & -0.2(4) \\
     O4  & $4b$ & 1/3 &2/3& 0.0077(20) & 0.7(3) \\
  \hline
  \hline
  \end{tabular}
 \end{center}
\end{table} 
%==========================================================================================================

\subsection{XRD}\label{sec:XRD}
First we use powder XRD to directly and quantitatively compare the refinements for the candidate polar and nonpolar structures. 
For sample preparation, a stoichiometric mixture of In$_2$O$_3$ (99.9\%) and Mn$_2$O$_3$ was placed in Au capsules and treated at 6~GPa 
in a belt-type high pressure apparatus at 1373~K for 30~min (heating rate 110~K/min). 
After heat treatment, the samples were quenched to room temperature, and the pressure was slowly released. 
The resultant samples were black dense pellets. Single-phase Mn$_2$O$_3$ was prepared from commercial MnO$_2$ (99.99\%) by heating in air at 923~K for 24~h.
The synchrotron XRD data were obtained on powdered samples at the BL02B2 beamline of SPring-8.\cite{NNuclInstrumMethodsPhysResSectA.467.1045}
They were collected in a $2\theta$ range from $2^{\circ}$ to $75^{\circ}$ with a step interval of 0.01 degrees and analyzed by the Rietveld method with RIETAN-2000.\cite{MaterSciForum.198.321}

An evaluation of the XRD data shown in Fig.~\ref{xrd} clearly reveals that the unit cell of InMnO$_3$ holds six formula units. 
This indicates a deviation of the $P6_3/mmc$ high-temperature phase due to unit-cell tripling with tilt-shear motions of the MnO$_5$ bipyramids as in the ferroelectric $R$MnO$_3$ compounds. 
The centrosymmetric and the noncentrosymmetric subgroups \textit{with the highest possible symmetry} 
that are compatible with a trimerization of the $P6_3/mmc$ high-temperature phase are $P6_3/mcm$ and $P6_3cm$, respectively. 
Here refinements clearly favor the latter structure which may contribute to former claims of the $P6_3cm$ symmetry for InMnO$_3$.\cite{InorgChem.50.3559}
However, in contrast to the case of the ferroelectric manganites, the structure refinements of the XRD data reveals equally good fits for the centrosymmetric space group $P\overline{3}c$ 
and the noncentrosymmetric space group $P6_3cm$~\cite{note-Rvalues} consistent with a previous observation.\cite{InorgChem.50.3559}
In Table~\ref{XRDTable} we report our refined atomic coordinates and lattice parameters within the $P\overline{3}c$ and $P6_3cm$ space groups. 
In general, structures refined with the correct space group are lower in energy than those refined with incorrect space groups.
Our DFT calculations for InMnO$_3$ at the atomic positions and cell parameters obtained in the two 
competing best-fit experimental refinements (calculation details given later) indicate that the nonpolar
$P\overline{3}c$ structure is $\sim$200~meV per formula unit (f.u.) lower in energy than the $P6_3cm$ structure. 
We therefore suggest the nonferroelectric $P\overline{3}c$ phase as the ground state for InMnO$_3$.

%==========================================================================================================
\begin{figure}
  \includegraphics[width=1\linewidth]{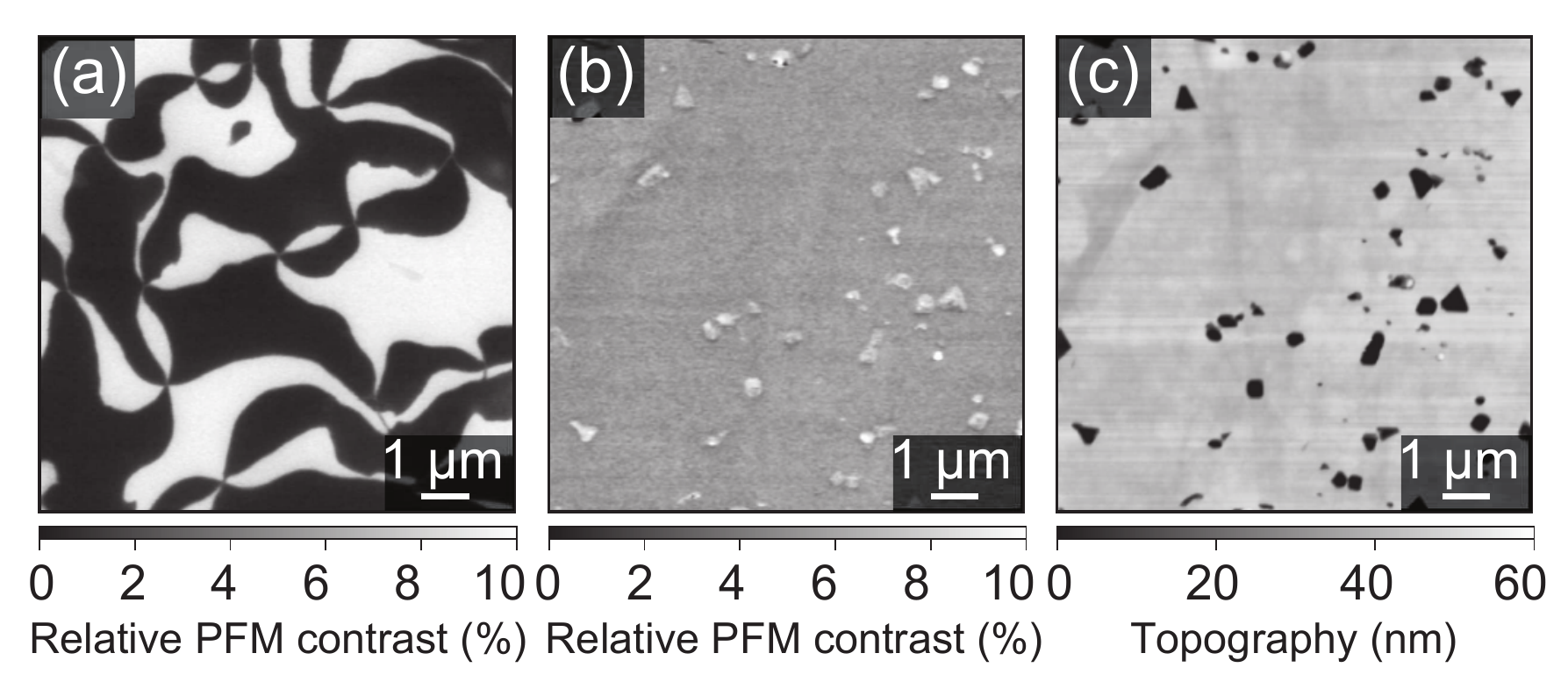}
  \caption{PFM images scaled relative to the response of periodically poled lithium niobate at room temperature, 
           for (a) YMnO$_3$ showing the known characteristic ferroelectric domain pattern, 
           and for (b) InMnO$_3$, where no ferroelectric domains are observed, and only topographical artifacts~(c) are visible.
}
  \label{pfm}
\end{figure}
%==========================================================================================================

\subsection{PFM}\label{sec:PFM}
To confirm this suggestion,
we next used PFM to probe directly for the presence of FE domains; this technique avoids ambiguities caused by sample 
leakiness which might have occurred in previous macroscopic polarization measurements.\cite{JApplPhys.100.076104}
In order to calibrate the response from InMnO$_3$, 
PFM and simultaneous scanning force microscopy (SFM) measurements with a commercial SFM (Solaris, NT-MDT) were carried out on YMnO$_3$ and InMnO$_3$.
All compounds for PFM and SHG measurements were grown by the flux method as $z$-oriented platelets.\cite{ActaCryst.16.957}
For PFM an ac-voltage of 14~V$_{\rm pp}$ at a frequency of ~40 kHz was applied to a conductive Pt-Ir coated probe (NSC 35, Mikromasch). 
The out-of-plane component of the piezoelectric response was recorded by the in-phase output channel of an external lock-in amplifier 
(SR830, Stanford Research) with a typical sensitivity of 200~$\mu$V and time constant of 10~ms. 
The PFM signal of each sample was normalized to a response of the $z$ face of PPLN ($d_{33}$=7~pm/V) in order to maintain comparability of the PFM
response, which was measured before and after each h-$R$MnO$_3$ measurement in order to exclude changes in the PFM sensitivity.\cite{note-symmetry}

Our results are summarized in the equally scaled Figs.~\ref{pfm}(a) and~\ref{pfm}(b). 
YMnO$_3$ reveals the familiar domain pattern of six intersecting AP+FE domains with alternating polarization $\pm P_z$.\cite{NatMat.9.253,ApplPhysLett.97.012904}
Strikingly, the InMnO$_3$ shows an almost homogeneous distribution of the PFM response with no sign of FE domains. 
The corresponding SFM data in Fig.~\ref{pfm}(c) show that the slightly brighter
speckles in the PFM image correspond to protrusions on the unpolished surface of the InMnO$_3$ sample. 
If one relates the contrast obtained for opposite domains in YMnO$_3$ to the spontaneous polarization of $5.6$~$\mu$C/cm$^2$, 
any polarization in InMnO$_3$ has to be at least two orders of magnitude smaller to avoid detection in our measurement. 
Our calculated polarization\cite{PhysRevB.47.1651} for the DFT-optimized $P6_3cm$ InMnO$_3$ structure 
is 4.8~$\mu$C/cm$^2$ which would certainly be detectable. 
% Oak reports 4.43~$\mu$C/cm$^2$ with LSDA and experimental lattice constants.
The absence of ferroelectricity (or of sub-resolution domains) in InMnO$_3$ is further supported by poling experiments with the SFM tip which did not induce any lasting change of the PFM response.

%==========================================================================================================
\begin{figure}
  \includegraphics[width=1\linewidth]{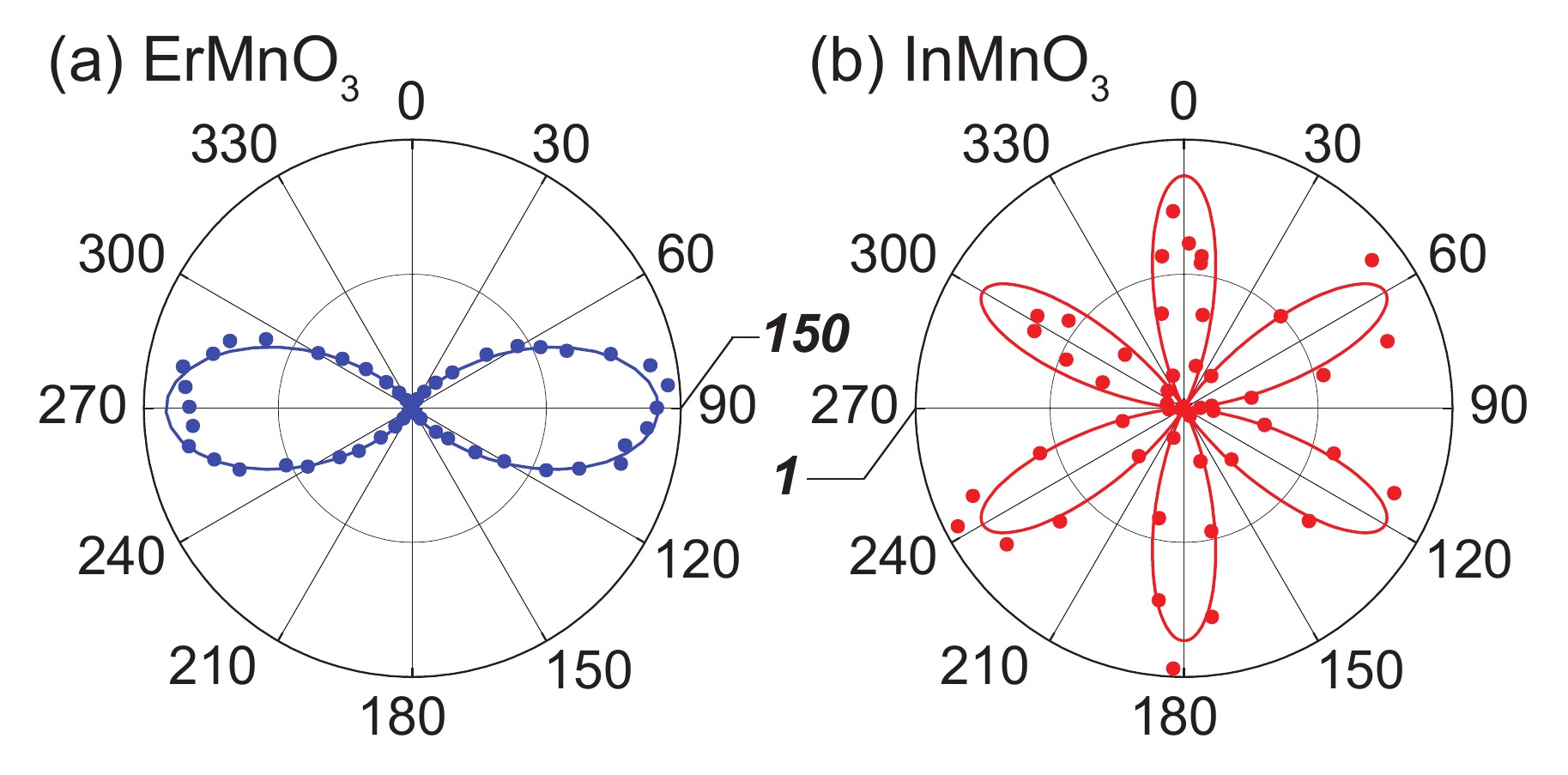}
  \caption{ SHG on (a) ErMnO$_3$ and (b) InMnO$_3$ at 5~K and 2.51~eV with light incident on the $z$-oriented surface of the single crystals along the [011] direction. 
           The anisotropy measurement was obtained by rotating the linear polarization of the incident light at $\omega$ and of the detected intensity at 2$\omega$ simultaneously by 360$^{\circ}$. 
           Lines are fits of the SHG data according to Ref.~\onlinecite{Book:Birss}.
           Scales in (a) and (b) differ by a factor 150.
           The two-fold SHG signal of ErMnO$_3$ indicates the presence of a spontaneous polarization whereas no indication for ferroelectricity in InMnO$_3$ was found. 
           Instead the six-fold pattern characteristic of the antiferromagnetic order in InMnO$_3$ is observed.
}
  \label{shg}
\end{figure}
%==========================================================================================================

\subsection{SHG}
Since PFM measurements could only be done under ambient conditions, we next used SHG to search for ferroelectric order at low temperature. 
As discussed in detail in Ref.~\onlinecite{JOptSocAmB.22.96}, the breaking of inversion symmetry by ferroelectric order leads to a characteristic SHG signal.  
The samples were mounted in a liquid-helium-operated cryostat and probed with 120 fs laser pulses in a standard transmission setup for SHG.\cite{JOptSocAmB.22.96}
For comparison, ErMnO$_3$ was chosen for the SHG data because, unlike YMnO$_3$, it has the same magnetic SHG spectrum as InMnO$_3$.
Figure~\ref{shg} shows the anisotropy of the SHG signal taken under identical conditions at 5~K on ferroelectric ErMnO$_3$ and on InMnO$_3$. 
The laser light was incident under 45$^{\circ}$ to the hexagonal crystal axes so that SHG components coupling to a spontaneous polarization along $z$ could be excited.\cite{JOptSocAmB.22.96}
ErMnO$_3$ shows the double lobe characteristic of the ferroelectric order.
In InMnO$_3$ the double lobe is absent. Instead a SHG signal with the sixfold anisotropy characteristic of the antiferromagnetic order and 150 times weaker intensity is found, indicating 
that the only order parameter is the antiferromagnetism of the Mn$^{3+}$ ions. 
We therefore conclude that InMnO$_3$ is not ferroelectric down to 5~K.

%==========================================================================================================
\begin{figure}
  \includegraphics[width=1\linewidth]{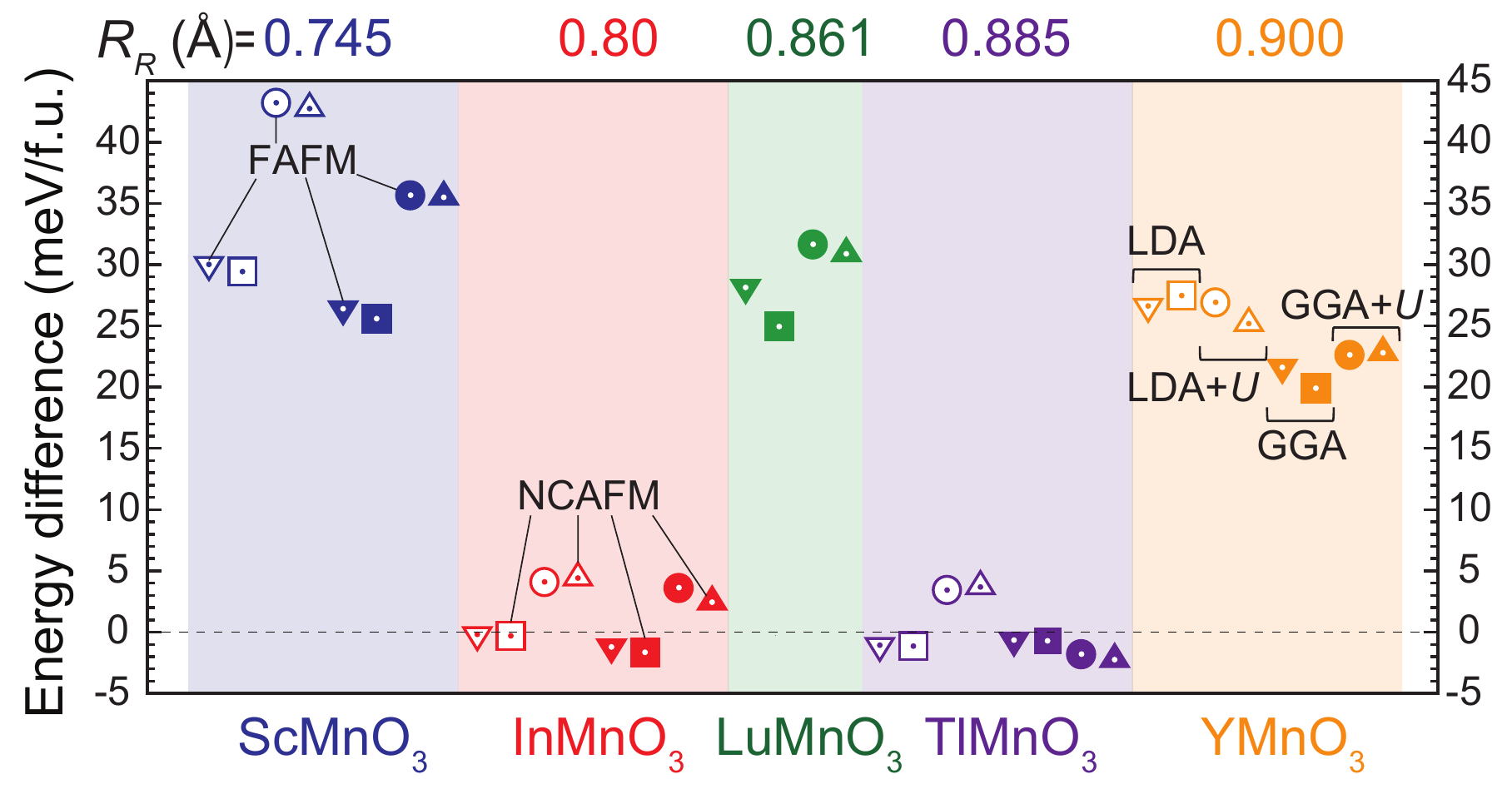}
  \caption{Calculated energy differences between the fully relaxed $P\overline{3}c$ phase and $P6_3cm$ phases, $\Delta E_{\rm str}$.
           Four exchange-correlation functionals, LDA, GGA, LDA+$U$, and GGA$+U$, and two spin configurations, FAFM and NCAFM (see text), were used.
           LDA/LDA+$U$ results for LuMnO$_3$ are not shown because {\sc vasp} does not provide an LDA Lu PAW potential.
           Positive $\Delta E_{\rm str}$ indicates that the ferroelectric $P6_3cm$ phase is stable over the $P\bar{3}c$ phase.
           The compounds are plotted in order of increasing $R$-site cation ionic radii, $R_R$ (Ref.~\onlinecite{ActaCryst.A32.751}).
}
  \label{energy_difference}
\end{figure}
%==========================================================================================================
\section{Theory}
To resolve the origin of the difference between InMnO$_3$ and the other hexagonal manganite h-$R$MnO$_3$ compounds, 
we used DFT calculations to evaluate the energy difference, $\Delta E_{\rm str}=E_{P\bar{3}c}-E_{P6_3cm}$, between the candidate $P6_3cm$ and $P\overline{3}c$ phases.
Our spin-polarized first principles calculations were performed using the projector augmented-wave (PAW) method\cite{PhysRevB.50.17953} as implemented in {\sc vasp}.\cite{PhysRevB.54.11169}
In this study, Sc 3$s$, 3$p$, 3$d$, and 4$s$, Y 4$s$, 4$p$, 4$d$, and 5$s$, In 5$s$ and 5$p$, Lu 5$p$, 5$d$, and 6$s$, Tl 6$s$ and 6$p$, Mn 3$d$ and 4$s$, and O 2$s$ and 2$p$ 
were described as valence electrons.
The PAW data set with radial cutoffs of 1.3, 1.4, 1.6, 1.6, 1.7, 1.2, and 0.8~$\AA$, respectively, for Sc, Y, In, Lu, Tl, Mn and O was employed.
The local density of states was also evaluated within the same spheres.
Wave functions were expanded with plane waves up to an energy cutoff of 500 eV.
All calculations were performed with 30-atom cells, which can describe unit cells of $P6_3cm$ and $P\overline{3}c$ phases. %\cite{note-phonon}
$k$-points were sampled with a $\Gamma$-centered 4$\times$4$\times$2 grid.
In addition to InMnO$_3$, we also calculated $\Delta E_{\rm str}$ for ScMnO$_3$, LuMnO$_3$, and YMnO$_3$, as well as for as-yet-unsynthesized TlMnO$_3$. 
To validate the results, we adopted four different exchange-correlation (XC) functionals:
local density approximation (LDA), generalized gradient approximation (GGA), LDA+$U$, and GGA+$U$,\cite{PhysRevB.23.5048,PhysRevLett.78.1396,PhysRevB.57.1505}
with the value for $U_{\rm eff}=U-J$ on the Mn-3$d$ orbitals set to 4~eV. In addition we tested
two different spin configurations, so-called frustrated antiferromagnetic (FAFM)\cite{JPhysCondensMatter.12.4947} 
and the noncollinear antiferromagnetic (NCAFM) adopted in Ref.~\onlinecite{PhysRevLett.106.047601}.
The lattice constants and internal positions were fully optimized in each case until the residual
stresses and forces converged to less than 0.1~GPa and 0.01~eV/\AA\ respectively. 

Figure~\ref{energy_difference} shows our calculated $\Delta E_{\rm str}$ values for the various functionals and magnetic configurations.
A positive $\Delta E_{\rm str}$ indicates that the ferroelectric structure is stable.
When the $R$ site is occupied with a ${\rm I\hspace{-.1em}I\hspace{-.1em}I}$b (Sc, Y, or Lu) ion, the $P6_3cm$ phase is more 
stable than the $P\overline{3}c$ phase consistent with the experimentally observed ferroelectricity. 
However, in the case of a ${\rm I\hspace{-.1em}I\hspace{-.1em}I}$a (In or Tl) ion, $\Delta E_{\rm str}$ is close to zero.
Note that in these calculations the lattice parameters and ionic positions for both structures were fully relaxed, within the constraint of the appropriate symmetry. 
[When we previously constrained our atomic positions and cell parameters to those obtained from the XRD analysis, 
the energy of the $P\overline{3}c$ phase is much lower ($\sim$200~meV/f.u.) than that of the $P6_3cm$ phase as we reported in Sec.~\ref{sec:XRD}.]
We note that $\Delta E_{\rm str}$ is quite insensitive to the magnetic configuration, but does show a dependence on
the choice of XC functional, and for InMnO$_3$, the {\it sign} of $\Delta E_{\rm str}$ depends on the functional, 
suggesting the possibility of competing low energy structures.\cite{note-hybrid}
[Previous calculations of the energy difference between $P6_3cm$ and $P\overline{3}c$ structures for InMnO$_3$ derived from XRD data (Sec.~\ref{sec:XRD}) 
and the polarization for InMnO$_3$ with the $P6_3cm$ symmetry (Sec.~\ref{sec:PFM}), and subsequent calculations use the GGA+$U$ and FAFM configuration.]

%==========================================================================================================
\begin{figure}
  \includegraphics[width=1.0\linewidth]{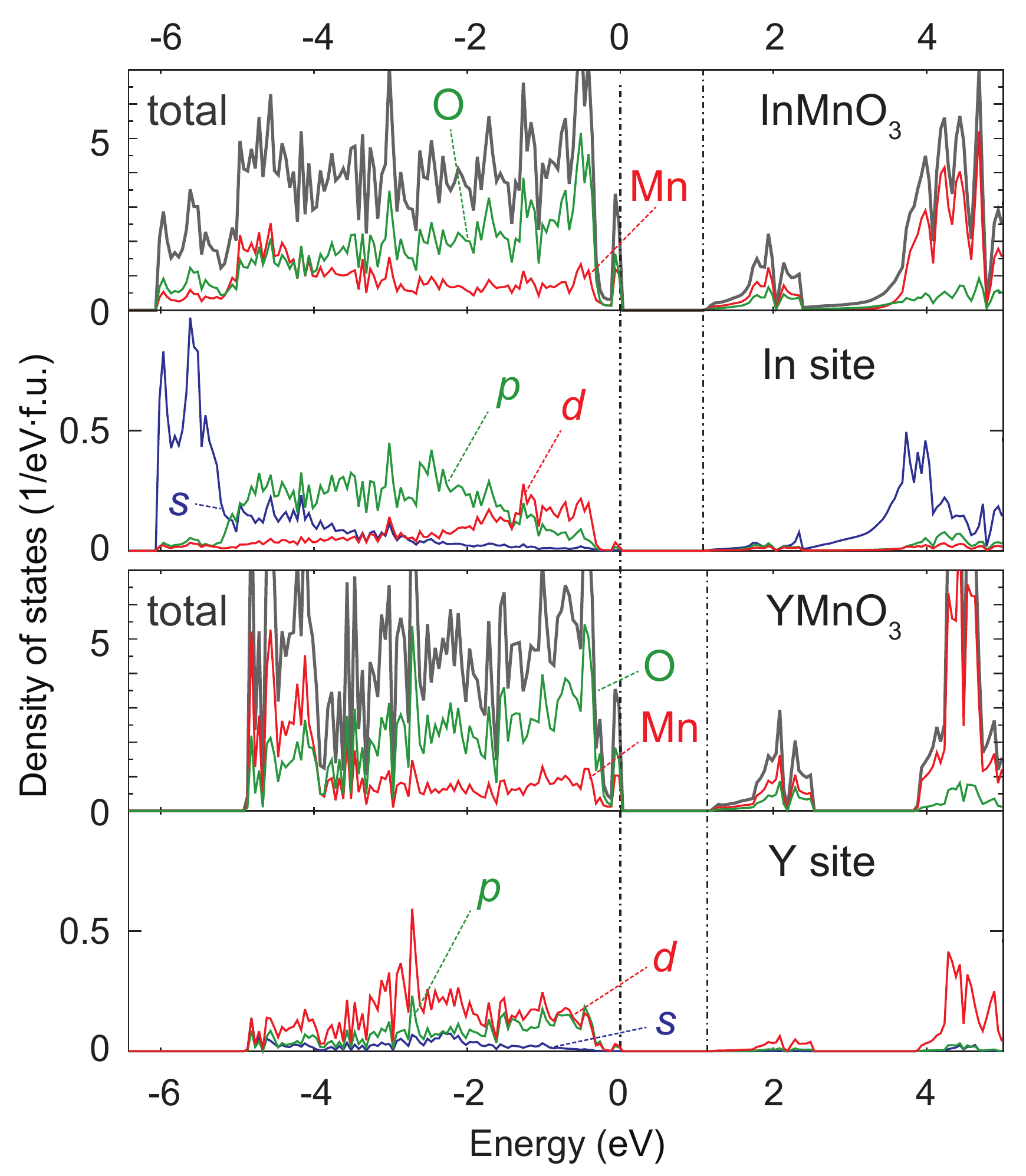}
  \caption{Calculated total and averaged local densities of states for InMnO$_3$ and YMnO$_3$ in the $P\overline{3}c$ phase.
           Zeroes of the horizontal axes are set to the top of the valence band.
           Valence bands are composed mainly of O-2$p$ and Mn-3$d$ orbitals.
           The difference of the electronic structures is mainly manifested at the formally unoccupied $R$ 5$s$ and 5$p$ local density of states.
}
\label{dos}
\end{figure}
% ==========================================================================================================

It is clear from Fig.~\ref{energy_difference} that $R_R$ is not the key factor in determining the phase stability.
Instead we focus on the different chemistry of the group ${\rm I\hspace{-.1em}I\hspace{-.1em}I}$a ions, compared to the 
group ${\rm I\hspace{-.1em}I\hspace{-.1em}I}$b ions.
It was previously suggested that the behavior of InMnO$_3$ is dominated by
high-lying occupied semicore $4d$ (In) electrons;\cite{PhysRevLett.106.047601} in contrast
the valence $d$ states are formally unoccupied in the  ${\rm I\hspace{-.1em}I\hspace{-.1em}I}$b ions. 
In fact it is well known that the presence or absence of semicore $d$ electrons can affect the structural stability
as illustrated by the different structures of MgO (rock salt) and ZnO (wurtzite, with semicore $d$s) in which the cations 
have very similar ionic radii ($r_{\rm MgO}=0.57~\AA$ and $r_{\rm ZnO}=0.60~\AA$ in four-coordination and $r_{\rm MgO}=0.72~\AA$ and $r_{\rm ZnO}=0.74~\AA$ in six-coordination).

In Fig.~\ref{dos}(a) we show our calculated densities of states (DOS)
for InMnO$_3$ and YMnO$_3$ (TlMnO$_3$, and ScMnO$_3$/LuMnO$_3$ behave analogously to InMnO$_3$ and YMnO$_3$, respectively),
both calculated within the $P\overline{3}c$ phase to allow a direct comparison.
In both cases, the valence bands consist mainly of Mn-$d$ (up-spin $e_{1g}$ and $e_{2g}$) and O-2$p$ states.
The main differences occur in the DOSs on the $R$ ions. 
The In ``semicore'' $4d$ states, however, form a narrow band that is around $-13$~eV below the top of
the valence band when the $d$ electrons are treated as valence (not shown). 
They do not directly contribute to covalent bonding with the oxygen anions,
in contrast to the suggestion in Ref.~\onlinecite{PhysRevLett.106.047601}.
The relevant difference is the substantially lower energy of the formally unoccupied $R$ $5s$ 
and $5p$ states in In compared with Y, caused by the well-known increase in nuclear charge without corresponding
increase in screening across the $4d$ series.
As a result, in InMnO$_3$ the $5s$ (and to a lesser extent $5p$) states, which would be completely empty
in the ionic limit, develop significant occupation through In-O$2p$ covalency, with occupied In $5s$ states
in fact forming the bottom of the valence band. 
(Similar behavior has been previously reported in other In oxides.\cite{JElectrMat.40.1501,PhysRevB.64.233111})
In YMnO$_3$, the Y $5s$ and $5p$ states are substantially higher in energy relative to the top of the valence
band and so their hybridization with O $2p$ and subsequent occupation is negligible. Instead there is a small
hybridization with the formally empty Y $4d$ states.
% The covalent bonds between $R$ and oxygen ions is also justified by decrease of the electrons at O$_{\rm ap}$ sites 
% (the averaged numbers of electrons within the oxygen PA sphere are 5.01 in InMnO$_3$ and 5.07 in YMnO$_3$).

\begin{figure}
  \includegraphics[width=0.9\linewidth]{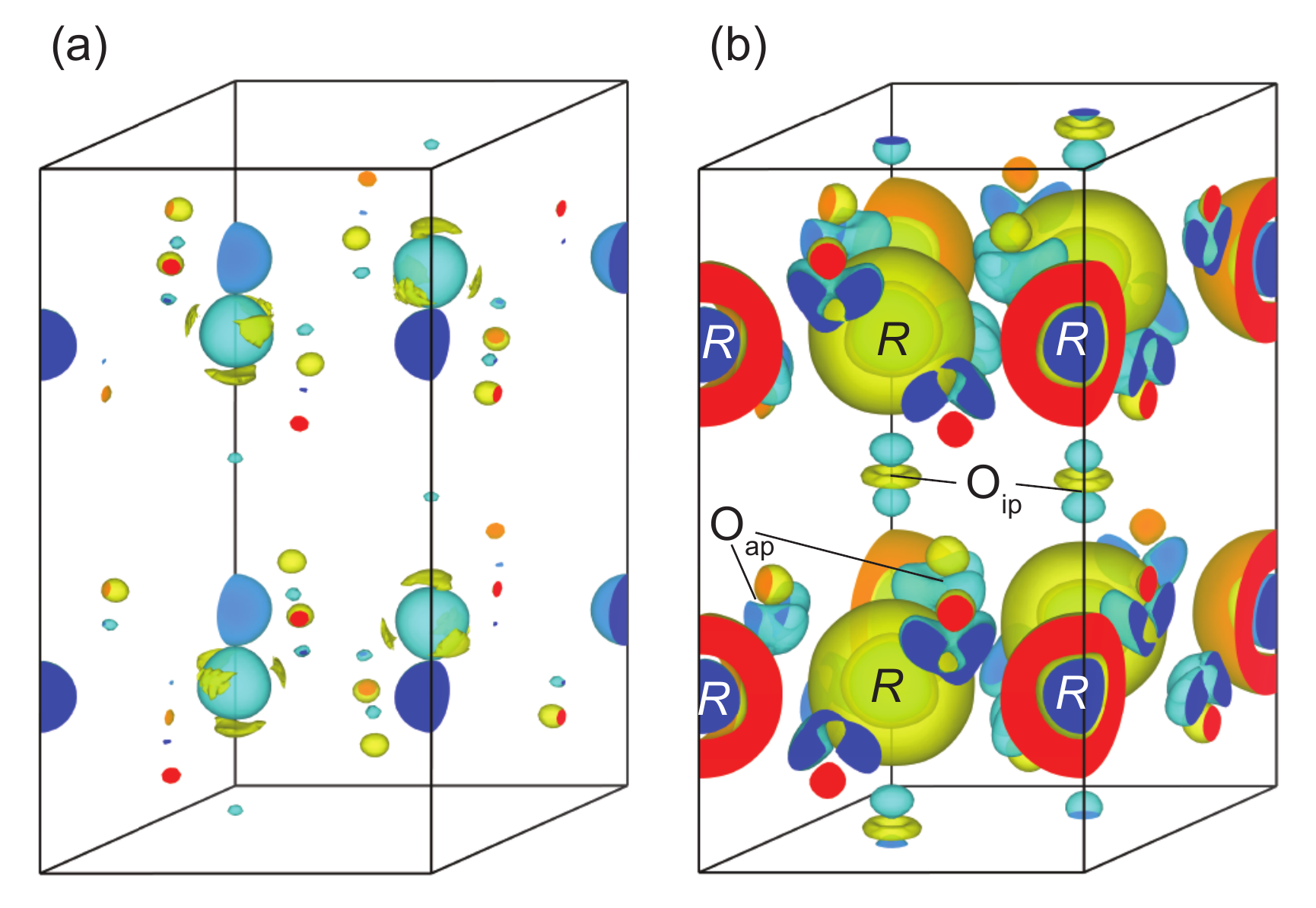}
  \caption{Charge density differences for valence band electrons between (a) LuMnO$_3$ and YMnO$_3$ and (b) InMnO$_3$ and YMnO$_3$ in the $P\overline{3}c$ phase.
           For the comparison, the same atomic positions were chosen for the compounds that are compared.
           The structures are constructed by averaging the DFT-optimized structures for (a) LuMnO$_3$ and YMnO$_3$, and (b) InMnO$_3$ and YMnO$_3$.
           Red and blue regions correspond to excess and deficiency of charge in LuMnO$_3$/InMnO$_3$ compared with YMnO$_3$.
           The yellow and light blue isosurfaces correspond to $0.03$ and $-0.03$ \AA$^{-3}$, respectively.
           Comparing with the charge density difference between LuMnO$_3$ and YMnO$_3$, InMnO$_3$ shows noticeable differences from YMnO$_3$ especially at the In site.
}
\label{charge-diff}
\end{figure}

The difference in covalency between InMnO$_3$ and YMnO$_3$ manifests particularly strikingly in the calculated 
valence charge densities at the $R$ sites. 
In Fig.~\ref{charge-diff} we show the valence charge density differences between LuMnO$_3$ and YMnO$_3$ and between InMnO$_3$ and YMnO$_3$ in the $P\overline{3}c$ structure. 
The charge density at the Mn sites is similar in all cases. 
Compared with YMnO$_3$ and LuMnO$_3$, however, InMnO$_3$, has a decrease in charge density at the O sites adjacent to the In ions 
and an increase at the outer region of the In site indicating charge transfer from oxygen to In and stronger In-O than Y-O or Lu-O 
covalent bond formation. 

Next we investigate how the additional covalency of the In-O and Tl-O bonds compared with those of Y-O and related compounds manifest
in the spring constants. 
Our calculated $z$-direction spring constants of the $R$ ions at the high-symmetry $2b$ sites in the $P\overline{3}c$ structure are 
2.7, 3.2, and 3.1 eV/$\AA^2$ for ScMnO$_3$, LuMnO$_3$, and YMnO$_3$, 
and are 4.4 and 4.2 eV/$\AA^2$ for InMnO$_3$ and TlMnO$_3$.
As expected, the strong In-O and Tl-O hybridization results in larger spring constants in the In and Tl compounds. 
The larger spring constants make the In and Tl ions reluctant to shift from their high-symmetry sites, favoring instead 
equal $R$-O$_{\rm ap}$ bond distances. This in turn favors the $P\overline{3}c$ structure, in which $\frac{1}{3}$ of the 
$R$ ions retain their fully 6-coordinated high-symmetry positions, over the ferroelectric $P6_3cm$ phase, in which all 
$R$ ions are displaced from the high-symmetry positions.

%==========================================================================================================
% ==========================================================================================================

We emphasize that the behavior here in which stronger covalency favors the nonferroelectric phase is completely 
different from that in conventional ferroelectrics such as BaTiO$_3$, in which stronger covalency favors the ferroelectric distortion through the second-order Jahn-Teller effect. 
In such conventional ferroelectrics, the Born effective charges $Z^* = \frac{\partial P}{\partial u}$ 
which participate actively in the re-hybridization are anomalously larger than the formal ionic charges,
reflecting the charge transfer that takes place during the ionic displacements to the ferroelectric phase; such
anomalous Born effective charges are signatures of instability toward a ferroelectric phase transition.\cite{PhysRevB.58.6224}
In both InMnO$_3$ and YMnO$_3$ the mechanism for the primary symmetry-lowering tilt distortion is geometric rather
than due to a rehybridization, and the Born effective charges on all atoms are nonanomalous.\cite{note-BEC}
In InMnO$_3$, the additional strong In-O covalency in the paraelectric phase {\it resists} the distortion of the 
In ions away from their high-symmetry positions favoring the $P\overline{3}c$ space group, whereas the lower 
Y-O covalency provides less resistance, allowing the additional Y-O displacements required to reach the $P6_3cm$ symmetry. 
In Ref.~\onlinecite{PhysRevLett.98.217601}, the hybridization between the Y-$3d$ and O-$2p$ orbitals was measured
using polarization-dependent x-ray absorption spectroscopy (XAS) at the O $K$-edge. 
Then the static charge occupancy in the Y $3d$ orbitals was equated with an anomalous dynamical Born effective charge, 
which led to the claim that this hybridization is responsible for the ferroelectricity in YMnO$_3$.
It is important to understand that the Born effective charge is the $derivative$ of the polarization with respect to ionic displacements, 
and is unrelated to the static orbital occupancy in a single structure: Partial hybridization of Y $3d$ with O $2p$, while 
clearly present both in the experiments and in earlier and subsequent first-principles calculations, is not indicative of an 
anomalous Born effective charge and therefore does not indicate tendency toward ferroelectricity. 

%==========================================================================================================
\begin{figure}
  \includegraphics[width=1.0\linewidth]{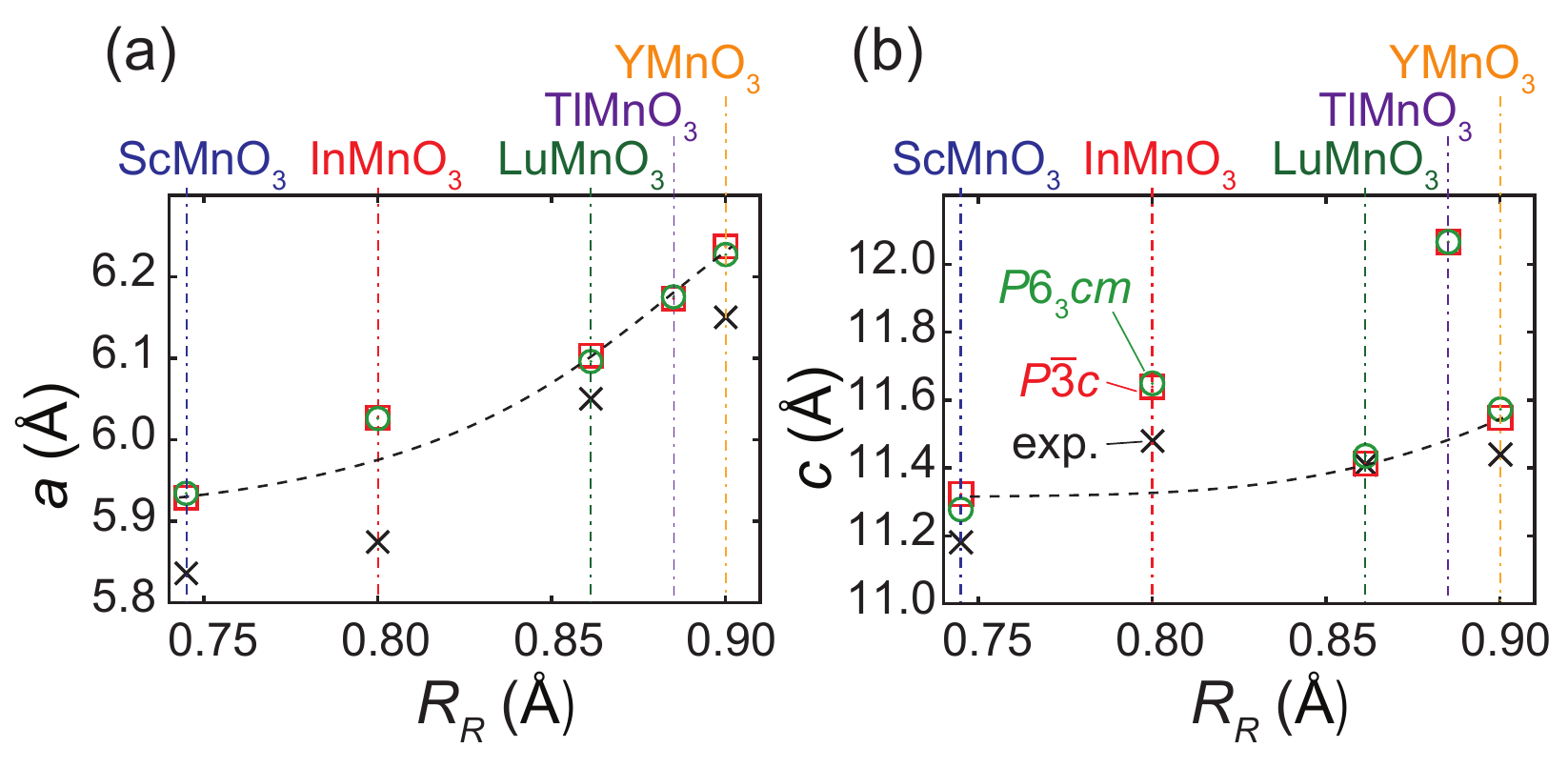}
  \caption{Calculated (a) in-plane and (b) out-of-plane lattice constants in the $P6_3cm$ ($\circ$) and $P\overline{3}c$ ($\Box$) phases.
           The experimental lattice constants are also shown ($\times$) (Ref.~\onlinecite{JSolidStateChem.143.132,JSolidStateChem.116.118,PhysRevB.83.094111}).
           The dashed lines indicate the trend of ${\rm I\hspace{-.1em}I\hspace{-.1em}I}$b manganites for a guide to eyes.}
  \label{structure_properties}
\end{figure}
%==========================================================================================================

Experimentally, it is known that InMnO$_3$ has an anomalously large $c$ lattice constant compared to the multiferroic hexagonal manganites.\cite{JSolidStateChem.116.118}
In Fig.~\ref{structure_properties} we plot our calculated lattice constants (with experimental values where available for comparison) 
of the manganites series as a function of ionic radii $R_R$.
We point out first that this is not a consequence of the different space group that we have established here;
our density functional calculations yield similar lattice constants for InMnO$_3$ in the $P\overline{3}c$ and $P6_3cm$ phases.
The calculated lattice constants are systematically overestimated compared with experiments as is typical of the GGA.
We see that the in-plane lattice constants, $a$, increase monotonically with $R_R$, with InMnO$_3$ showing only a small calculated anomaly.
In contrast, the $c$ lattice constant of InMnO$_3$ deviates strongly from the trend shown by the 
${\rm I\hspace{-.1em}I\hspace{-.1em}I}$b manganites, both in our calculations and in experiment.
Since this deviation is also identified in TlMnO$_3$, the anomalously large $c$ likely originates from covalency in ${\rm I\hspace{-.1em}I\hspace{-.1em}I}$a manganites as shown in Fig.~\ref{charge-diff}.

\section{Discussion}
Although the centrosymmetric $P\overline{3}c$ phase that we propose in this work for InMnO$_3$ may be seemingly less 
attractive compared with the ferroelectric $P6_3cm$ structure, 
our results have implications for the multiferroic hexagonal manganites as a whole.
Since the tilt pattern of the YMnO$_3$ structure subsequently allows for the development of ferroelectricity, 
whereas that of the InMnO$_3$ structure does not, the subtle chemical bonding differences identified here that favor one 
tilt pattern over another in turn determine whether the resulting structure can be multiferroic. 
Specifically, we have discussed here that an {\it absence} of $R$-O hybridization is required to favor the YMnO$_3$ tilt pattern over the InMnO$_3$ 
tilt pattern; an absence of $R$-O hybridization is therefore a requirement for ferroelectricity in the hexagonal manganites. 
Earlier theoretical papers correctly noted the electrostatic origin of the ``geometric ferroelectricity'' mechanism 
in the hexagonal manganites;\cite{NatMat.430.541} we now understand that the relative stability
of ferroelectric and nonferroelectric structures is determined by a competition between electrostatics (favoring the ferroelectric phase) and covalency (favoring the nonferroelectric phase). 

%==========================================================================================================
\begin{figure}
  \includegraphics[width=1.0\linewidth]{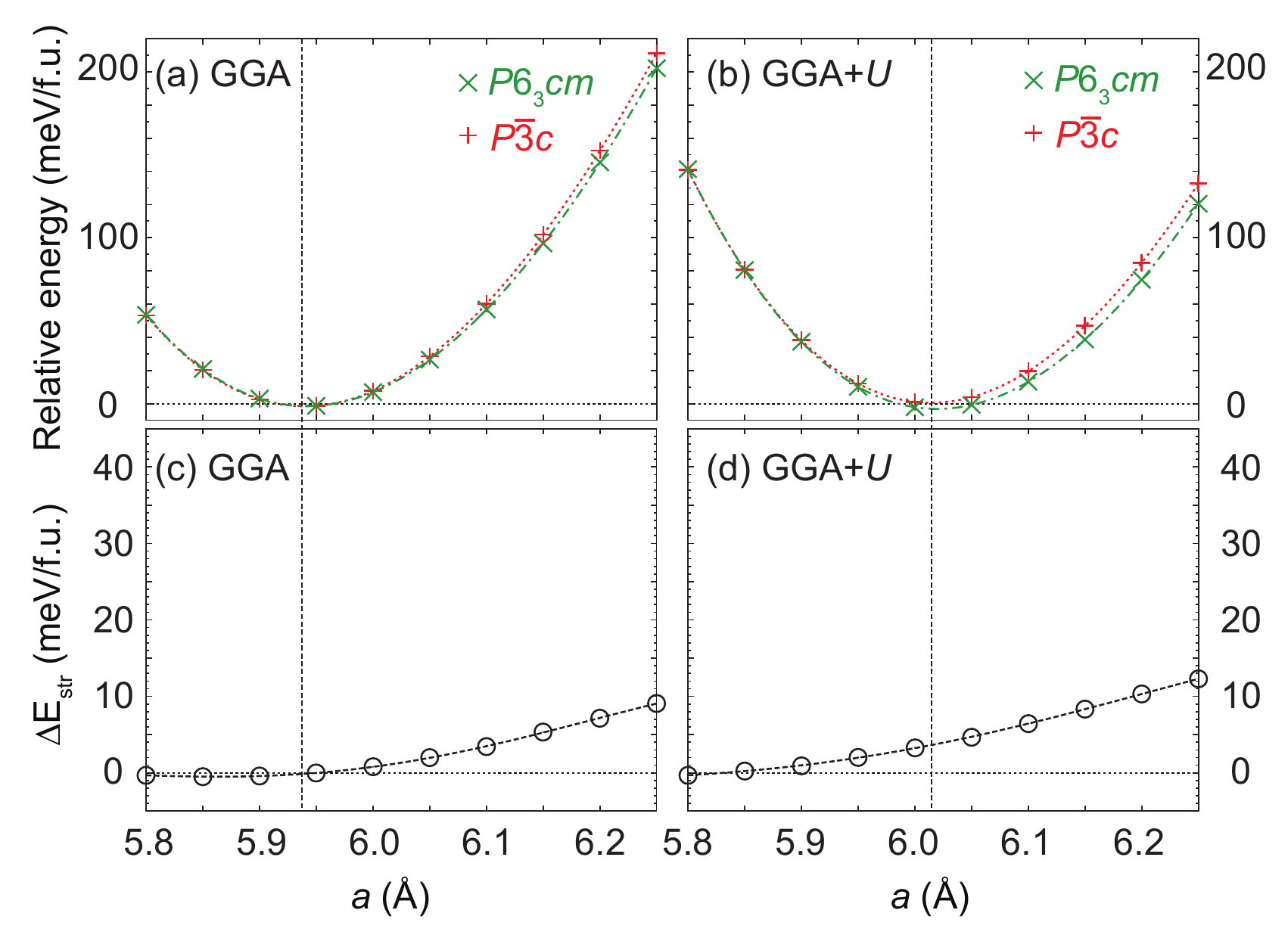}
  \caption{(a),(b) Calculated InMnO$_3$ $P\overline{3}c$ and $P6_3cm$ structure energies relative to the equilibrium $P\overline{3}c$ phase
           as a function of in-plane lattice constant $a$ by the GGA and GGA$+U$.
           (c),(d) Energy differences $\Delta E_{\rm str}$ between the $P\overline{3}c$ and $P6_3cm$ phases. 
           Positive values indicate that the ferroelectric $P6_3cm$ phase is stable over the $P\overline{3}c$ phase.
           The fitted lines were obtained with fourth-order polynomials.
           The vertical dashed lines indicate the calculated equilibrium lattice constants of InMnO$_3$ $P\overline{3}c$ phase.}
  \label{epitaxial_strain}
\end{figure}
%==========================================================================================================

In addition, since the polar $P6_3cm$ and nonpolar $P\overline{3}c$ phases are close in energy in InMnO$_3$ as shown in Fig.~\ref{energy_difference}, 
it might be expected that their relative stability could be changed using external perturbations such as epitaxial strain.
Figure~\ref{epitaxial_strain} shows the calculated energies of the $P6_3cm$ and $P\overline{3}c$ phases and their energy differences as a function of in-plane lattice constant $a$, with the out-of-plane lattice constant $c$ and internal positions fully relaxed for each $a$ value.
Because the critical strain of the phase boundary depends on the $U$ value, we performed the calculations using both the GGA and GGA$+U$ methods.
We see that the $\Delta E_{\rm str}$ increases with the increasing in-plane lattice constant, indicating a larger in-plane lattice constant could develop the polar $P6_3cm$.
Interestingly, the ferroelectric polarization develops in the out-of-plane direction, in striking contrast to the behavior in perovskites.\cite{PhysRevB.72.144101}
Therefore we anticipate that InMnO$_3$ could be tuned into the polar $P6_3cm$ structure using tensile strain.

Finally we mention that a recent transmission electron microscopy study of the domain walls
in ferroelectric hexagonal TmMnO$_3$ and LuMnO$_3$\cite{PhysRevB.85.020102} revealed that a domain wall 
structure at the edges of the sample is similar to the nonpolar $P\overline{3}c$ InMnO$_3$ structure: The $R$ ion at the wall is at the centrosymmetric
position, with one neighbor displaced in the up-direction and one in the down-direction. Detailed calculations of the domain-wall structure in $R$MnO$_3$ are ongoing. 

\section{Conclusion}
In summary, we have proposed a different nonferroelectric ground state structure in the hexagonal manganite
InMnO$_3$, and we predict its occurrence in as-yet-unsynthesized hexagonal TlMnO$_3$. 
The proposed phase has $P\overline{3}c$ symmetry, and is closely related to the usual $P6_3cm$ ferroelectric ground state but with a different 
pattern of polyhedral tilts that retains the center of inversion. The energy balance between the two
related phases is determined by a competition between electrostatics and $R$-O covalency, with $lower$
$R$-O covalency favoring the ferroelectric structure.
Thus, the {\it absence} of ferroelectricity in InMnO$_3$ reveals to us the reason for the {\it presence} of ferroelectricity (and therefore multiferroicity) in the other h-$R$MnO$_3$ compounds.

\begin{acknowledgments}
We thank M. Bieringer, Department of Chemistry, University of Manitoba for providing the InMnO$_3$ single crystals.
Y.K. acknowledges support by JSPS Postdoctoral Fellowships for Research Abroad.
Y.K., M.L., N.L., M.F., and N.A.S. acknowledge support from ETH Zurich, and A.A.B. acknowledges support from MANA WPI Initiative (MEXT, Japan), FIRST Program (JSPS), and JSPS Grant No. 22246083. 
The SXRD was performed under Proposals No. 2009A1136 and No. 2010A1215.
The visualization of crystal structures and charge density differences were performed with {\sc vesta}.\cite{JApplCryst.41.653}
\end{acknowledgments}

%\bibliography{InMnO3}

\begin{thebibliography}{44}
\expandafter\ifx\csname natexlab\endcsname\relax\def\natexlab#1{#1}\fi
\expandafter\ifx\csname bibnamefont\endcsname\relax
  \def\bibnamefont#1{#1}\fi
\expandafter\ifx\csname bibfnamefont\endcsname\relax
  \def\bibfnamefont#1{#1}\fi
\expandafter\ifx\csname citenamefont\endcsname\relax
  \def\citenamefont#1{#1}\fi
\expandafter\ifx\csname url\endcsname\relax
  \def\url#1{\texttt{#1}}\fi
\expandafter\ifx\csname urlprefix\endcsname\relax\def\urlprefix{URL }\fi
\providecommand{\bibinfo}[2]{#2}
\providecommand{\eprint}[2][]{\url{#2}}

\bibitem[{\citenamefont{Choi et~al.}(2010)\citenamefont{Choi, Horibe, Yi, Choi,
  Wu, and Cheong}}]{NatMat.9.253}
\bibinfo{author}{\bibfnamefont{T.}~\bibnamefont{Choi}},
  \bibinfo{author}{\bibfnamefont{Y.}~\bibnamefont{Horibe}},
  \bibinfo{author}{\bibfnamefont{H.}~\bibnamefont{Yi}},
  \bibinfo{author}{\bibfnamefont{Y.}~\bibnamefont{Choi}},
  \bibinfo{author}{\bibfnamefont{W.}~\bibnamefont{Wu}}, \bibnamefont{and}
  \bibinfo{author}{\bibfnamefont{S.-W.} \bibnamefont{Cheong}},
  \bibinfo{journal}{Nature Mater.} \textbf{\bibinfo{volume}{9}},
  \bibinfo{pages}{253} (\bibinfo{year}{2010}).

\bibitem[{\citenamefont{Jungk et~al.}(2010)\citenamefont{Jungk, Hoffmann,
  Fiebig, and Soergel}}]{ApplPhysLett.97.012904}
\bibinfo{author}{\bibfnamefont{T.}~\bibnamefont{Jungk}},
  \bibinfo{author}{\bibfnamefont{A.}~\bibnamefont{Hoffmann}},
  \bibinfo{author}{\bibfnamefont{M.}~\bibnamefont{Fiebig}}, \bibnamefont{and}
  \bibinfo{author}{\bibfnamefont{E.}~\bibnamefont{Soergel}},
  \bibinfo{journal}{Appl. Phys. Lett.} \textbf{\bibinfo{volume}{97}},
  \bibinfo{pages}{012904} (\bibinfo{year}{2010}).

\bibitem[{\citenamefont{Meier et~al.}(2012)\citenamefont{Meier, Seidel, Cano,
  Delaney, Kumagai, Mostovoy, Spaldin, Ramesh, and Fiebig}}]{NatMat.submitted}
\bibinfo{author}{\bibfnamefont{D.}~\bibnamefont{Meier}},
  \bibinfo{author}{\bibfnamefont{J.}~\bibnamefont{Seidel}},
  \bibinfo{author}{\bibfnamefont{A.}~\bibnamefont{Cano}},
  \bibinfo{author}{\bibfnamefont{K.}~\bibnamefont{Delaney}},
  \bibinfo{author}{\bibfnamefont{Y.}~\bibnamefont{Kumagai}},
  \bibinfo{author}{\bibfnamefont{M.}~\bibnamefont{Mostovoy}},
  \bibinfo{author}{\bibfnamefont{N.}~\bibnamefont{Spaldin}},
  \bibinfo{author}{\bibfnamefont{R.}~\bibnamefont{Ramesh}}, \bibnamefont{and}
  \bibinfo{author}{\bibfnamefont{M.}~\bibnamefont{Fiebig}},
  \bibinfo{journal}{Nature Mater.} \textbf{\bibinfo{volume}{11}},
  \bibinfo{pages}{284} (\bibinfo{year}{2012}).

\bibitem[{\citenamefont{Van~Aken et~al.}(2004)\citenamefont{Van~Aken, Palstra,
  Filippetti, and Spaldin}}]{NatMat.430.541}
\bibinfo{author}{\bibfnamefont{B.}~\bibnamefont{Van~Aken}},
  \bibinfo{author}{\bibfnamefont{T.}~\bibnamefont{Palstra}},
  \bibinfo{author}{\bibfnamefont{A.}~\bibnamefont{Filippetti}},
  \bibnamefont{and} \bibinfo{author}{\bibfnamefont{N.}~\bibnamefont{Spaldin}},
  \bibinfo{journal}{Nature Mater.} \textbf{\bibinfo{volume}{3}},
  \bibinfo{pages}{164} (\bibinfo{year}{2004}).

\bibitem[{\citenamefont{Fennie and Rabe}(2005)}]{PhysRevB.72.100103}
\bibinfo{author}{\bibfnamefont{C.~J.} \bibnamefont{Fennie}} \bibnamefont{and}
  \bibinfo{author}{\bibfnamefont{K.~M.} \bibnamefont{Rabe}},
  \bibinfo{journal}{Phys. Rev. B} \textbf{\bibinfo{volume}{72}},
  \bibinfo{pages}{100103} (\bibinfo{year}{2005}).

\bibitem[{\citenamefont{Greedan et~al.}(1995)\citenamefont{Greedan, Bieringer,
  Britten, Giaquinta, and zur Loye}}]{JSolidStateChem.116.118}
\bibinfo{author}{\bibfnamefont{J.~E.} \bibnamefont{Greedan}},
  \bibinfo{author}{\bibfnamefont{M.}~\bibnamefont{Bieringer}},
  \bibinfo{author}{\bibfnamefont{J.~F.} \bibnamefont{Britten}},
  \bibinfo{author}{\bibfnamefont{D.~M.} \bibnamefont{Giaquinta}},
  \bibnamefont{and} \bibinfo{author}{\bibfnamefont{H.~C.} \bibnamefont{zur
  Loye}}, \bibinfo{journal}{J. Solid State Chem.}
  \textbf{\bibinfo{volume}{116}}, \bibinfo{pages}{118} (\bibinfo{year}{1995}).

\bibitem[{\citenamefont{Belik et~al.}(2009)\citenamefont{Belik, Kamba, Savinov,
  Nuzhnyy, Tachibana, Takayama-Muromachi, and Goian}}]{PhysRevB.79.054411}
\bibinfo{author}{\bibfnamefont{A.~A.} \bibnamefont{Belik}},
  \bibinfo{author}{\bibfnamefont{S.}~\bibnamefont{Kamba}},
  \bibinfo{author}{\bibfnamefont{M.}~\bibnamefont{Savinov}},
  \bibinfo{author}{\bibfnamefont{D.}~\bibnamefont{Nuzhnyy}},
  \bibinfo{author}{\bibfnamefont{M.}~\bibnamefont{Tachibana}},
  \bibinfo{author}{\bibfnamefont{E.}~\bibnamefont{Takayama-Muromachi}},
  \bibnamefont{and} \bibinfo{author}{\bibfnamefont{V.}~\bibnamefont{Goian}},
  \bibinfo{journal}{Phys. Rev. B} \textbf{\bibinfo{volume}{79}},
  \bibinfo{pages}{054411} (\bibinfo{year}{2009}).

\bibitem[{\citenamefont{Fabr\`eges et~al.}(2011)\citenamefont{Fabr\`eges,
  Mirebeau, Petit, Bonville, and Belik}}]{PhysRevB.84.054455}
\bibinfo{author}{\bibfnamefont{X.}~\bibnamefont{Fabr\`eges}},
  \bibinfo{author}{\bibfnamefont{I.}~\bibnamefont{Mirebeau}},
  \bibinfo{author}{\bibfnamefont{S.}~\bibnamefont{Petit}},
  \bibinfo{author}{\bibfnamefont{P.}~\bibnamefont{Bonville}}, \bibnamefont{and}
  \bibinfo{author}{\bibfnamefont{A.~A.} \bibnamefont{Belik}},
  \bibinfo{journal}{Phys. Rev. B} \textbf{\bibinfo{volume}{84}},
  \bibinfo{pages}{054455} (\bibinfo{year}{2011}).

\bibitem[{\citenamefont{Rusakov et~al.}(2011)\citenamefont{Rusakov, Belik,
  Kamba, Savinov, Nuzhnyy, Kolodiazhnyi, Yamaura, Takayama-Muromachi,
  Borodavka, and Kroupa}}]{InorgChem.50.3559}
\bibinfo{author}{\bibfnamefont{D.~A.} \bibnamefont{Rusakov}},
  \bibinfo{author}{\bibfnamefont{A.~A.} \bibnamefont{Belik}},
  \bibinfo{author}{\bibfnamefont{S.}~\bibnamefont{Kamba}},
  \bibinfo{author}{\bibfnamefont{M.}~\bibnamefont{Savinov}},
  \bibinfo{author}{\bibfnamefont{D.}~\bibnamefont{Nuzhnyy}},
  \bibinfo{author}{\bibfnamefont{T.}~\bibnamefont{Kolodiazhnyi}},
  \bibinfo{author}{\bibfnamefont{K.}~\bibnamefont{Yamaura}},
  \bibinfo{author}{\bibfnamefont{E.}~\bibnamefont{Takayama-Muromachi}},
  \bibinfo{author}{\bibfnamefont{F.}~\bibnamefont{Borodavka}},
  \bibnamefont{and} \bibinfo{author}{\bibfnamefont{J.}~\bibnamefont{Kroupa}},
  \bibinfo{journal}{Inorg. Chem.} \textbf{\bibinfo{volume}{50}},
  \bibinfo{pages}{3559} (\bibinfo{year}{2011}).

\bibitem[{\citenamefont{Serrao et~al.}(2006)\citenamefont{Serrao, Krupanidhi,
  Bhattacharjee, Waghmare, Kundu, and Rao}}]{JApplPhys.100.076104}
\bibinfo{author}{\bibfnamefont{C.~R.} \bibnamefont{Serrao}},
  \bibinfo{author}{\bibfnamefont{S.~B.} \bibnamefont{Krupanidhi}},
  \bibinfo{author}{\bibfnamefont{J.}~\bibnamefont{Bhattacharjee}},
  \bibinfo{author}{\bibfnamefont{U.~V.} \bibnamefont{Waghmare}},
  \bibinfo{author}{\bibfnamefont{A.~K.} \bibnamefont{Kundu}}, \bibnamefont{and}
  \bibinfo{author}{\bibfnamefont{C.~N.~R.} \bibnamefont{Rao}},
  \bibinfo{journal}{J. Appl. Phys.} \textbf{\bibinfo{volume}{100}},
  \bibinfo{pages}{076104} (\bibinfo{year}{2006}).

\bibitem[{\citenamefont{Scott}(2008)}]{JPhysCondensMatter.20.021001}
\bibinfo{author}{\bibfnamefont{J.~F.} \bibnamefont{Scott}},
  \bibinfo{journal}{J. Phys.: Condens. Matter} \textbf{\bibinfo{volume}{20}},
  \bibinfo{pages}{021001} (\bibinfo{year}{2008}).

\bibitem[{\citenamefont{Nishibori et~al.}(2001)\citenamefont{Nishibori, Takata,
  Kato, Sakata, Kubota, Aoyagi, Kuroiwa, Yamakata, and
  M.~Ikeda}}]{NNuclInstrumMethodsPhysResSectA.467.1045}
\bibinfo{author}{\bibfnamefont{E.}~\bibnamefont{Nishibori}},
  \bibinfo{author}{\bibfnamefont{M.}~\bibnamefont{Takata}},
  \bibinfo{author}{\bibfnamefont{K.}~\bibnamefont{Kato}},
  \bibinfo{author}{\bibfnamefont{M.}~\bibnamefont{Sakata}},
  \bibinfo{author}{\bibfnamefont{Y.}~\bibnamefont{Kubota}},
  \bibinfo{author}{\bibfnamefont{S.}~\bibnamefont{Aoyagi}},
  \bibinfo{author}{\bibfnamefont{Y.}~\bibnamefont{Kuroiwa}},
  \bibinfo{author}{\bibfnamefont{M.}~\bibnamefont{Yamakata}}, \bibnamefont{and}
  \bibinfo{author}{\bibfnamefont{N.}~\bibnamefont{M.~Ikeda}},
  \bibinfo{journal}{N. Nucl. Instrum. Methods Phys. Res. Sect. A}
  \textbf{\bibinfo{volume}{467}}, \bibinfo{pages}{1045} (\bibinfo{year}{2001}).

\bibitem[{\citenamefont{Izumi and Ikeda}(2000)}]{MaterSciForum.198.321}
\bibinfo{author}{\bibfnamefont{F.}~\bibnamefont{Izumi}} \bibnamefont{and}
  \bibinfo{author}{\bibfnamefont{T.}~\bibnamefont{Ikeda}},
  \bibinfo{journal}{Mater. Sci. Forum} \textbf{\bibinfo{volume}{198}},
  \bibinfo{pages}{321} (\bibinfo{year}{2000}).

\bibitem[{not({\natexlab{a}})}]{note-Rvalues}
\bibinfo{note}{The $R$ values for nonpolar $P\overline{3}c$ are $R_{\rm
  wp}$=6.61\%, $R_{\rm p}$=4.90\%, $R_{\rm I}$=2.48\%, and $R_{\rm F}$=1.39\%.
  Those for polar $P6_3cm$ phase are $R_{\rm wp}$=6.64\%, $R_{\rm p}$=4.94\%,
  $R_{\rm I}$=2.36\%, and $R_{\rm F}$=1.33\%. Those for nonpolar $P6_3/mcm$
  are $R_{\rm wp}$=12.94\%, $R_{\rm p}$=8.90\%, $R_{\rm I}$=6.08\%, and $R_{\rm
  F}$=3.34\%.}

\bibitem[{\citenamefont{Yakel et~al.}(1963)\citenamefont{Yakel, Koehler,
  Bertaut, and Forrat}}]{ActaCryst.16.957}
\bibinfo{author}{\bibfnamefont{H.~L.} \bibnamefont{Yakel}},
  \bibinfo{author}{\bibfnamefont{W.~C.} \bibnamefont{Koehler}},
  \bibinfo{author}{\bibfnamefont{E.~F.} \bibnamefont{Bertaut}},
  \bibnamefont{and} \bibinfo{author}{\bibfnamefont{E.~F.}
  \bibnamefont{Forrat}}, \bibinfo{journal}{Acta Cryst.}
  \textbf{\bibinfo{volume}{16}}, \bibinfo{pages}{957} (\bibinfo{year}{1963}).

\bibitem[{not({\natexlab{b}})}]{note-symmetry}
\bibinfo{note}{Although the $P\overline{3}c$ space group belongs to the
  trigonal group, in this study we use the term ``hexagonal manganite'' for simplicity to
  describe the structure class with $P\overline{3}c$ and $P6_3cm$ space groups
  for simplicity.}

\bibitem[{\citenamefont{King-Smith and Vanderbilt}(1993)}]{PhysRevB.47.1651}
\bibinfo{author}{\bibfnamefont{R.~D.} \bibnamefont{King-Smith}}
  \bibnamefont{and}
  \bibinfo{author}{\bibfnamefont{D.}~\bibnamefont{Vanderbilt}},
  \bibinfo{journal}{Phys. Rev. B} \textbf{\bibinfo{volume}{47}},
  \bibinfo{pages}{1651} (\bibinfo{year}{1993}).

\bibitem[{\citenamefont{Birss}(1966)}]{Book:Birss}
\bibinfo{author}{\bibfnamefont{R.}~\bibnamefont{Birss}},
  \emph{\bibinfo{title}{Symmetry and Magnetism}}
  (\bibinfo{publisher}{North-Holland}, \bibinfo{address}{Amsterdam},
  \bibinfo{year}{1966}).

\bibitem[{\citenamefont{Fiebig et~al.}(2005)\citenamefont{Fiebig, Pavlov, and
  Pisarev}}]{JOptSocAmB.22.96}
\bibinfo{author}{\bibfnamefont{M.}~\bibnamefont{Fiebig}},
  \bibinfo{author}{\bibfnamefont{V.}~\bibnamefont{Pavlov}}, \bibnamefont{and}
  \bibinfo{author}{\bibfnamefont{R.}~\bibnamefont{Pisarev}},
  \bibinfo{journal}{J. Opt. Soc. Am. B} \textbf{\bibinfo{volume}{22}},
  \bibinfo{pages}{96} (\bibinfo{year}{2005}).

\bibitem[{\citenamefont{Shannon}(1976)}]{ActaCryst.A32.751}
\bibinfo{author}{\bibfnamefont{R.~D.} \bibnamefont{Shannon}},
  \bibinfo{journal}{Acta Cryst.} \textbf{\bibinfo{volume}{A32}},
  \bibinfo{pages}{751} (\bibinfo{year}{1976}).

\bibitem[{\citenamefont{Bl$\ddot{\rm o}$chl}(1994)}]{PhysRevB.50.17953}
\bibinfo{author}{\bibfnamefont{P.~E.} \bibnamefont{Bl$\ddot{\rm o}$chl}},
  \bibinfo{journal}{Phys. Rev. B} \textbf{\bibinfo{volume}{50}},
  \bibinfo{pages}{17953} (\bibinfo{year}{1994}).

\bibitem[{\citenamefont{Kresse and Furthm$\ddot{\rm
  u}$ller}(1996)}]{PhysRevB.54.11169}
\bibinfo{author}{\bibfnamefont{G.}~\bibnamefont{Kresse}} \bibnamefont{and}
  \bibinfo{author}{\bibfnamefont{J.}~\bibnamefont{Furthm$\ddot{\rm u}$ller}},
  \bibinfo{journal}{Phys. Rev. B} \textbf{\bibinfo{volume}{54}},
  \bibinfo{pages}{11169} (\bibinfo{year}{1996}).

\bibitem[{\citenamefont{Perdew and Zunger}(1981)}]{PhysRevB.23.5048}
\bibinfo{author}{\bibfnamefont{J.~P.} \bibnamefont{Perdew}} \bibnamefont{and}
  \bibinfo{author}{\bibfnamefont{A.}~\bibnamefont{Zunger}},
  \bibinfo{journal}{Phys. Rev. B} \textbf{\bibinfo{volume}{23}},
  \bibinfo{pages}{5048} (\bibinfo{year}{1981}).

\bibitem[{\citenamefont{Perdew et~al.}(1997)\citenamefont{Perdew, Burke, and
  Ernzerhof}}]{PhysRevLett.78.1396}
\bibinfo{author}{\bibfnamefont{J.~P.} \bibnamefont{Perdew}},
  \bibinfo{author}{\bibfnamefont{K.}~\bibnamefont{Burke}}, \bibnamefont{and}
  \bibinfo{author}{\bibfnamefont{M.}~\bibnamefont{Ernzerhof}},
  \bibinfo{journal}{Phys. Rev. Lett.} \textbf{\bibinfo{volume}{78}},
  \bibinfo{pages}{1396} (\bibinfo{year}{1997}).

\bibitem[{\citenamefont{Dudarev et~al.}(1998)\citenamefont{Dudarev, Botton,
  Savrasov, Humphreys, and Sutton}}]{PhysRevB.57.1505}
\bibinfo{author}{\bibfnamefont{S.~L.} \bibnamefont{Dudarev}},
  \bibinfo{author}{\bibfnamefont{G.~A.} \bibnamefont{Botton}},
  \bibinfo{author}{\bibfnamefont{S.~Y.} \bibnamefont{Savrasov}},
  \bibinfo{author}{\bibfnamefont{C.~J.} \bibnamefont{Humphreys}},
  \bibnamefont{and} \bibinfo{author}{\bibfnamefont{A.~P.}
  \bibnamefont{Sutton}}, \bibinfo{journal}{Phys. Rev. B}
  \textbf{\bibinfo{volume}{57}}, \bibinfo{pages}{1505} (\bibinfo{year}{1998}).

\bibitem[{\citenamefont{Medvedeva et~al.}(2000)\citenamefont{Medvedeva,
  Mryasov, Korotin, Anisimov, and Freeman}}]{JPhysCondensMatter.12.4947}
\bibinfo{author}{\bibfnamefont{J.}~\bibnamefont{Medvedeva}},
  \bibinfo{author}{\bibfnamefont{O.}~\bibnamefont{Mryasov}},
  \bibinfo{author}{\bibfnamefont{M.}~\bibnamefont{Korotin}},
  \bibinfo{author}{\bibfnamefont{V.}~\bibnamefont{Anisimov}}, \bibnamefont{and}
  \bibinfo{author}{\bibfnamefont{A.}~\bibnamefont{Freeman}},
  \bibinfo{journal}{J. Phys.: Condens. Matter} \textbf{\bibinfo{volume}{12}},
  \bibinfo{pages}{4947} (\bibinfo{year}{2000}).

\bibitem[{\citenamefont{Oak et~al.}(2011)\citenamefont{Oak, Lee, Jang, Goh,
  Choi, and Scott}}]{PhysRevLett.106.047601}
\bibinfo{author}{\bibfnamefont{M.-A.} \bibnamefont{Oak}},
  \bibinfo{author}{\bibfnamefont{J.-H.} \bibnamefont{Lee}},
  \bibinfo{author}{\bibfnamefont{H.~M.} \bibnamefont{Jang}},
  \bibinfo{author}{\bibfnamefont{J.~S.} \bibnamefont{Goh}},
  \bibinfo{author}{\bibfnamefont{H.~J.} \bibnamefont{Choi}}, \bibnamefont{and}
  \bibinfo{author}{\bibfnamefont{J.~F.} \bibnamefont{Scott}},
  \bibinfo{journal}{Phys. Rev. Lett.} \textbf{\bibinfo{volume}{106}},
  \bibinfo{pages}{047601} (\bibinfo{year}{2011}).

\bibitem[{not({\natexlab{c}})}]{note-hybrid}
\bibinfo{note}{We also performed calculations with the HSE06 hybrid
  functional (Ref.~\onlinecite{JChemPhys.124.219906}) that is known to describe the
  electronic structure for 3$d$ transition metal compounds more
  precisely (Ref.~\onlinecite{PhysRevB.85.033203,PhysRevB.83.214421,InorgChem.51.4560,NewJPhys.12.093026,PhysRevB.85.054417}).
  The calculated energy differences with FAFM configuration and
  3$\times$3$\times$2 $k$-points are 0.6 meV/f.u. for InMnO$_3$ and 19.9
  meV/f.u. for YMnO$_3$, both of which are similar to the results in
  Fig.~\ref{energy_difference}}.

\bibitem[{\citenamefont{Zhang and Wang}(2011)}]{JElectrMat.40.1501}
\bibinfo{author}{\bibfnamefont{Y.}~\bibnamefont{Zhang}} \bibnamefont{and}
  \bibinfo{author}{\bibfnamefont{Y.}~\bibnamefont{Wang}}, \bibinfo{journal}{J.
  Electr. Mat.} \textbf{\bibinfo{volume}{40}}, \bibinfo{pages}{1501}
  (\bibinfo{year}{2011}).

\bibitem[{\citenamefont{Mryasov and Freeman}(2001)}]{PhysRevB.64.233111}
\bibinfo{author}{\bibfnamefont{O.~N.} \bibnamefont{Mryasov}} \bibnamefont{and}
  \bibinfo{author}{\bibfnamefont{A.~J.} \bibnamefont{Freeman}},
  \bibinfo{journal}{Phys. Rev. B} \textbf{\bibinfo{volume}{64}},
  \bibinfo{pages}{233111} (\bibinfo{year}{2001}).

\bibitem[{\citenamefont{Ghosez et~al.}(1998)\citenamefont{Ghosez, Michenaud,
  and Gonze}}]{PhysRevB.58.6224}
\bibinfo{author}{\bibfnamefont{P.}~\bibnamefont{Ghosez}},
  \bibinfo{author}{\bibfnamefont{J.-P.} \bibnamefont{Michenaud}},
  \bibnamefont{and} \bibinfo{author}{\bibfnamefont{X.}~\bibnamefont{Gonze}},
  \bibinfo{journal}{Phys. Rev. B} \textbf{\bibinfo{volume}{58}},
  \bibinfo{pages}{6224} (\bibinfo{year}{1998}).

\bibitem[{not({\natexlab{d}})}]{note-BEC}
\bibinfo{note}{We calculated $Z^*_R$ by displacing $R$ cations in $P6_3/mmc$ in
  the $z$-direction. The obtained $Z^*_R$ for InMnO$_3$ and YMnO$_3$ are
  3.8$\left|e\right|$ and 4.1$\left|e\right|$. They are slightly higher than
  the formal charges, 3$\left|e\right|$, but smaller than the $Z^*$s on $d^0$
  cations in perovskites such as BaTiO$_3$ ($Z^*_{\rm
  Ti}=7.3\left|e\right|$) (Ref.~\onlinecite{PhysRevB.58.6224}}).

\bibitem[{\citenamefont{Cho et~al.}(2007)\citenamefont{Cho, Kim, Park, Rho,
  Park, Noh, Kim, Oh, Park, Ahn et~al.}}]{PhysRevLett.98.217601}
\bibinfo{author}{\bibfnamefont{D.-Y.} \bibnamefont{Cho}},
  \bibinfo{author}{\bibfnamefont{J.-Y.} \bibnamefont{Kim}},
  \bibinfo{author}{\bibfnamefont{B.-G.} \bibnamefont{Park}},
  \bibinfo{author}{\bibfnamefont{K.-J.} \bibnamefont{Rho}},
  \bibinfo{author}{\bibfnamefont{J.-H.} \bibnamefont{Park}},
  \bibinfo{author}{\bibfnamefont{H.-J.} \bibnamefont{Noh}},
  \bibinfo{author}{\bibfnamefont{B.}~\bibnamefont{Kim}},
  \bibinfo{author}{\bibfnamefont{S.-J.} \bibnamefont{Oh}},
  \bibinfo{author}{\bibfnamefont{H.-M.} \bibnamefont{Park}},
  \bibinfo{author}{\bibfnamefont{J.-S.} \bibnamefont{Ahn}},
  \bibnamefont{$et$~$al$.}, \bibinfo{journal}{Phys. Rev. Lett.}
  \textbf{\bibinfo{volume}{98}}, \bibinfo{pages}{217601}
  (\bibinfo{year}{2007}).

\bibitem[{\citenamefont{Bieringer and Greedan}(1999)}]{JSolidStateChem.143.132}
\bibinfo{author}{\bibfnamefont{M.}~\bibnamefont{Bieringer}} \bibnamefont{and}
  \bibinfo{author}{\bibfnamefont{J.}~\bibnamefont{Greedan}},
  \bibinfo{journal}{J. Solid State Chem.} \textbf{\bibinfo{volume}{143}},
  \bibinfo{pages}{132} (\bibinfo{year}{1999}).

\bibitem[{\citenamefont{Gibbs et~al.}(2011)\citenamefont{Gibbs, Knight, and
  Lightfoot}}]{PhysRevB.83.094111}
\bibinfo{author}{\bibfnamefont{A.~S.} \bibnamefont{Gibbs}},
  \bibinfo{author}{\bibfnamefont{K.~S.} \bibnamefont{Knight}},
  \bibnamefont{and}
  \bibinfo{author}{\bibfnamefont{P.}~\bibnamefont{Lightfoot}},
  \bibinfo{journal}{Phys. Rev. B} \textbf{\bibinfo{volume}{83}},
  \bibinfo{pages}{094111} (\bibinfo{year}{2011}).

\bibitem[{\citenamefont{Di\'eguez et~al.}(2005)\citenamefont{Di\'eguez, Rabe,
  and Vanderbilt}}]{PhysRevB.72.144101}
\bibinfo{author}{\bibfnamefont{O.}~\bibnamefont{Di\'eguez}},
  \bibinfo{author}{\bibfnamefont{K.~M.} \bibnamefont{Rabe}}, \bibnamefont{and}
  \bibinfo{author}{\bibfnamefont{D.}~\bibnamefont{Vanderbilt}},
  \bibinfo{journal}{Phys. Rev. B} \textbf{\bibinfo{volume}{72}},
  \bibinfo{pages}{144101} (\bibinfo{year}{2005}).

\bibitem[{\citenamefont{Zhang et~al.}(2012)\citenamefont{Zhang, Wang, Wei, Yu,
  Gu, Hirata, Chen, Jin, Yao, Wang et~al.}}]{PhysRevB.85.020102}
\bibinfo{author}{\bibfnamefont{Q.~H.} \bibnamefont{Zhang}},
  \bibinfo{author}{\bibfnamefont{L.~J.} \bibnamefont{Wang}},
  \bibinfo{author}{\bibfnamefont{X.~K.} \bibnamefont{Wei}},
  \bibinfo{author}{\bibfnamefont{R.~C.} \bibnamefont{Yu}},
  \bibinfo{author}{\bibfnamefont{L.}~\bibnamefont{Gu}},
  \bibinfo{author}{\bibfnamefont{A.}~\bibnamefont{Hirata}},
  \bibinfo{author}{\bibfnamefont{M.~W.} \bibnamefont{Chen}},
  \bibinfo{author}{\bibfnamefont{C.~Q.} \bibnamefont{Jin}},
  \bibinfo{author}{\bibfnamefont{Y.}~\bibnamefont{Yao}},
  \bibinfo{author}{\bibfnamefont{Y.~G.} \bibnamefont{Wang}},
  \bibnamefont{$et$~$al$.}, \bibinfo{journal}{Phys. Rev. B}
  \textbf{\bibinfo{volume}{85}}, \bibinfo{pages}{020102}
  (\bibinfo{year}{2012}).

\bibitem[{\citenamefont{Momma and Izumi}(2008)}]{JApplCryst.41.653}
\bibinfo{author}{\bibfnamefont{K.}~\bibnamefont{Momma}} \bibnamefont{and}
  \bibinfo{author}{\bibfnamefont{F.}~\bibnamefont{Izumi}}, \bibinfo{journal}{J.
  Appl. Cryst.} \textbf{\bibinfo{volume}{41}}, \bibinfo{pages}{653}
  (\bibinfo{year}{2008}).

\bibitem[{\citenamefont{Heyd et~al.}(2006)\citenamefont{Heyd, Scuseria, and
  Ernzerhof}}]{JChemPhys.124.219906}
\bibinfo{author}{\bibfnamefont{J.}~\bibnamefont{Heyd}},
  \bibinfo{author}{\bibfnamefont{G.~E.} \bibnamefont{Scuseria}},
  \bibnamefont{and}
  \bibinfo{author}{\bibfnamefont{M.}~\bibnamefont{Ernzerhof}},
  \bibinfo{journal}{J. Chem. Phys.} \textbf{\bibinfo{volume}{124}},
  \bibinfo{pages}{219906} (\bibinfo{year}{2006}).

\bibitem[{\citenamefont{Kumagai et~al.}(2012)\citenamefont{Kumagai, Soda, Oba,
  Seko, and Tanaka}}]{PhysRevB.85.033203}
\bibinfo{author}{\bibfnamefont{Y.}~\bibnamefont{Kumagai}},
  \bibinfo{author}{\bibfnamefont{Y.}~\bibnamefont{Soda}},
  \bibinfo{author}{\bibfnamefont{F.}~\bibnamefont{Oba}},
  \bibinfo{author}{\bibfnamefont{A.}~\bibnamefont{Seko}}, \bibnamefont{and}
  \bibinfo{author}{\bibfnamefont{I.}~\bibnamefont{Tanaka}},
  \bibinfo{journal}{Phys. Rev. B} \textbf{\bibinfo{volume}{85}},
  \bibinfo{pages}{033203} (\bibinfo{year}{2012}).

\bibitem[{\citenamefont{Akamatsu et~al.}(2011)\citenamefont{Akamatsu, Kumagai,
  Oba, Fujita, Murakami, Tanaka, and Tanaka}}]{PhysRevB.83.214421}
\bibinfo{author}{\bibfnamefont{H.}~\bibnamefont{Akamatsu}},
  \bibinfo{author}{\bibfnamefont{Y.}~\bibnamefont{Kumagai}},
  \bibinfo{author}{\bibfnamefont{F.}~\bibnamefont{Oba}},
  \bibinfo{author}{\bibfnamefont{K.}~\bibnamefont{Fujita}},
  \bibinfo{author}{\bibfnamefont{H.}~\bibnamefont{Murakami}},
  \bibinfo{author}{\bibfnamefont{K.}~\bibnamefont{Tanaka}}, \bibnamefont{and}
  \bibinfo{author}{\bibfnamefont{I.}~\bibnamefont{Tanaka}},
  \bibinfo{journal}{Phys. Rev. B} \textbf{\bibinfo{volume}{83}},
  \bibinfo{pages}{214421} (\bibinfo{year}{2011}).

\bibitem[{\citenamefont{Akamatsu et~al.}(2012)\citenamefont{Akamatsu, Fujita,
  Hayashi, Kawamoto, Kumagai, Zong, Iwata, Oba, Tanaka, and
  Tanaka}}]{InorgChem.51.4560}
\bibinfo{author}{\bibfnamefont{H.}~\bibnamefont{Akamatsu}},
  \bibinfo{author}{\bibfnamefont{K.}~\bibnamefont{Fujita}},
  \bibinfo{author}{\bibfnamefont{H.}~\bibnamefont{Hayashi}},
  \bibinfo{author}{\bibfnamefont{T.}~\bibnamefont{Kawamoto}},
  \bibinfo{author}{\bibfnamefont{Y.}~\bibnamefont{Kumagai}},
  \bibinfo{author}{\bibfnamefont{Y.}~\bibnamefont{Zong}},
  \bibinfo{author}{\bibfnamefont{K.}~\bibnamefont{Iwata}},
  \bibinfo{author}{\bibfnamefont{F.}~\bibnamefont{Oba}},
  \bibinfo{author}{\bibfnamefont{I.}~\bibnamefont{Tanaka}}, \bibnamefont{and}
  \bibinfo{author}{\bibfnamefont{K.}~\bibnamefont{Tanaka}},
  \bibinfo{journal}{Inorg. Chem.} \textbf{\bibinfo{volume}{51}},
  \bibinfo{pages}{4560} (\bibinfo{year}{2012}).

\bibitem[{\citenamefont{Stroppa et~al.}(2010)\citenamefont{Stroppa, Marsman,
  Kresse, and Picozzi}}]{NewJPhys.12.093026}
\bibinfo{author}{\bibfnamefont{A.}~\bibnamefont{Stroppa}},
  \bibinfo{author}{\bibfnamefont{M.}~\bibnamefont{Marsman}},
  \bibinfo{author}{\bibfnamefont{G.}~\bibnamefont{Kresse}}, \bibnamefont{and}
  \bibinfo{author}{\bibfnamefont{S.}~\bibnamefont{Picozzi}},
  \bibinfo{journal}{New J. Phys.} \textbf{\bibinfo{volume}{12}},
  \bibinfo{pages}{093026} (\bibinfo{year}{2010}).

\bibitem[{\citenamefont{Hong et~al.}(2012)\citenamefont{Hong, Stroppa,
  \'I\~niguez, Picozzi, and Vanderbilt}}]{PhysRevB.85.054417}
\bibinfo{author}{\bibfnamefont{J.}~\bibnamefont{Hong}},
  \bibinfo{author}{\bibfnamefont{A.}~\bibnamefont{Stroppa}},
  \bibinfo{author}{\bibfnamefont{J.}~\bibnamefont{\'I\~niguez}},
  \bibinfo{author}{\bibfnamefont{S.}~\bibnamefont{Picozzi}}, \bibnamefont{and}
  \bibinfo{author}{\bibfnamefont{D.}~\bibnamefont{Vanderbilt}},
  \bibinfo{journal}{Phys. Rev. B} \textbf{\bibinfo{volume}{85}},
  \bibinfo{pages}{054417} (\bibinfo{year}{2012}).

\end{thebibliography}

\end{document}